\documentclass[9pt,shortpaper,twoside,web]{IEEEtran}
\usepackage{cite}
\usepackage{amsmath,amssymb,amsfonts}
\usepackage{algorithmic}
\usepackage{graphicx}
\usepackage{textcomp}

\usepackage{mathtools}
\usepackage{amsthm}

\usepackage{booktabs}

\usepackage{siunitx}

\usepackage[acronym, style=alttree, toc=true, shortcuts, xindy, nomain, nonumberlist]{glossaries}

\usepackage{cleveref}

\usepackage{hyphenat}
\usepackage{bm}

\usepackage{subcaption}
\usepackage{tikz}
\usetikzlibrary{positioning,arrows,fit,backgrounds}

\usepackage{pgfplots}
\pgfplotsset{compat=newest}
\usetikzlibrary{plotmarks}
\usetikzlibrary{arrows.meta}
\usetikzlibrary{positioning} 
\usetikzlibrary{shapes,arrows}
\usepgfplotslibrary{patchplots}
\newlength\fsize
\newlength\fwidth
\newlength\fheight

\newcommand{\bs}[1]{\boldsymbol{#1}}
\newcommand{\ol}[1]{\overline{#1}}

\newcommand{\norm}[1]{\left\lVert#1\right\rVert} 
\newcommand{\setdeff}[2]{
	\left\{
		\ifblank{#1}{}{#1 \hspace{.1cm} \middle| \hspace{.1cm}}
		#2
	\right\}
} 

\newcommand{\lrsum}[1]{\left(#1\right)} 
\newcommand{\intl}[2]{ \mathbb{I}_{{#1},{#2}} } 

\newcommand{\lr}[1]{\left(#1\right)} 

\DeclarePairedDelimiterXPP\onenorm[1]{}\lVert\rVert{_1}{\ifblank{#1}{\:\cdot\:}{#1}} 
\DeclarePairedDelimiterXPP\twonorm[1]{}\lVert\rVert{_2}{\ifblank{#1}{\:\cdot\:}{#1}} 

\newtheorem{theorem}{Theorem}
\newtheorem{assumption}{Assumption}
\newtheorem{remark}{Remark}
\newtheorem{definition}{Definition}
\newtheorem{lemma}{Lemma}

\newtheorem{problem}{Problem}

\newglossary{symbols}{sym}{sbl}{List of Symbols}
\newglossary{notation}{not}{nt}{Notation}

\glsaddkey{dimension}{\glsentrytext{\glslabel}}{\glsentryd}{\GLsentryd}{\glsd}{\Glsd}{\GLSd}
\glsaddkey{body}{\glsentrytext{\glslabel}}{\glsentrybody}{\GLsentrybody}{\glsbody}{\Glsbody}{\GLSbody}

\newacronym{ocp}{OCP}{optimal control problem}
\newacronym{MPC}{MPC}{Model Predictive Control}
\newacronym{cvpm}{CVPM-MPC}{CVPM-MPC}
\newacronym{CVPM}{CVPM}{constraint violation probability minimization}
\newacronym{RMPC}{RMPC}{Robust Model Predictive Control}
\newacronym{SMPC}{SMPC}{Stochastic Model Predictive Control}
\newacronym{SCMPC}{SCMPC}{Scenario Model Predictive Control}
\newacronym{cbf}{CBF}{Control Barrier Function}
\newacronym{ISS}{ISS}{input-to-state stable}

\newglossaryentry{c1}{type=symbols,
	sort={c1},
	name={the safe case},
	plural={safe cases},
	description={case 1},
}
\newglossaryentry{c1b}{type=symbols,
	sort={c1},
	name={safe case},
	plural={safe cases},
	description={case 1},
}
\newglossaryentry{C1}{type=symbols,
	sort={c1},
	name={The safe case},
	plural={Safe Cases},
	description={case 1},
}
\newglossaryentry{CC1}{type=symbols,
	sort={c1},
	name={Safe Case},
	plural={Safe Cases},
	description={case 1},
}
\newglossaryentry{c2}{type=symbols,
	sort={c2},
	name={the probabilistic case},
	plural={probabilistic cases},
	description={case 2},
}
\newglossaryentry{c2b}{type=symbols,
	sort={c2},
	name={probabilistic case},
	plural={probabilistic cases},
	description={case 2},
}
\newglossaryentry{C2}{type=symbols,
	sort={c2},
	name={The probabilistic case},
	plural={Probabilistic Cases},
	description={case 2},
}
\newglossaryentry{CC2}{type=symbols,
	sort={c1},
	name={Probabilistic Case},
	plural={Safe Cases},
	description={case 1},
}
\newglossaryentry{tCC2}{type=symbols,
	sort={c1},
	name={the Probabilistic Case},
	plural={Safe Cases},
	description={case 1},
}

\newglossaryentry{Vf}{type=symbols,
	sort={u1},
	name={\ensuremath{V_{\text{f}}}},
	plural={\ensuremath{\bs{u}}},
	description={Input vector with the dimension \gls{dimu} at time $k$},
}

\newglossaryentry{nx}{type=symbols,
	sort={u1},
	name={\ensuremath{\ol{\bm{x}}}},
	plural={\ensuremath{\bs{u}}},
	description={Input vector with the dimension \gls{dimu} at time $k$},
}

\newglossaryentry{Xrci}{type=symbols,
	sort={u1},
	name={\ensuremath{\mathcal{X}_{\text{rpi}}}},
	plural={\ensuremath{\mathcal{X}_{\text{rpi}}}},
	description={Robust control invariant set},
}

\newglossaryentry{input}{type=symbols,
	sort={u1},
	name={\ensuremath{\bs{u}_t}},
	plural={\ensuremath{\bs{u}}},
	description={Input vector with the dimension \gls{dimu} at time $k$},
}
\newglossaryentry{state}{type=symbols,
	sort={x1},
	name={\ensuremath{\bs{x}_t}},
	plural={\ensuremath{\bs{x}}},
	description={State vector with the dimension \gls{dimx} at time $k$}
}

\newglossaryentry{error}{type=symbols,
	sort={error},
	name={\ensuremath{\bs{e}_t}},
	plural={\ensuremath{\bs{e}}},
	description={Error between State vector and $\xi$}
}

\newglossaryentry{dist}{type=symbols,
	sort={w1},
	name={\ensuremath{\bs{w}_t}},
	plural={\ensuremath{\bs{w}}},
	description={Disturbance vector with the dimension \gls{dimw} at time $k$}
}
\newglossaryentry{nomInput}{type=symbols,
	sort={u2},
	name={\ensuremath{\bar{\bs{u}}_t}},
	plural={\ensuremath{\bar{\bs{u}}}},
	description={Nominal/mean vector of the input vector},
}
\newglossaryentry{nomState}{type=symbols,
	sort={x2},
	name={\ensuremath{\bar{\bs{x}}_t}},
	plural={\ensuremath{\bar{\bs{x}}}},
	description={Nominal/mean vector of the state vector},
}
\newglossaryentry{inputs}{type=symbols,
	sort={u3},
	name={\ensuremath{\underline{\bs{u}}_k}},
	plural={\ensuremath{\underline{\bs{u}}}},
	description={Stacked vector with all inputs  $\glspl{input}_0$ to $\glspl{input}_i$}
}
\newglossaryentry{states}{type=symbols,
	sort={x3},
	name={\ensuremath{\underline{\bs{x}}_k}},
	plural={\ensuremath{\underline{\bs{x}}}},
	description={Stacked vector with states $\glspl{state}_1$ to $\glspl{state}_i$}
}
\newglossaryentry{dists}{type=symbols,
	sort={w3},
	name={\ensuremath{\underline{\bs{w}}_k}},
	plural={\ensuremath{\underline{\bs{w}}}},
	description={Stacked vector with inputs $\glspl{dist}_0$ to $\glspl{dist}_i$}
}
\newglossaryentry{nomInputs}{type=symbols,
	sort={u4},
	name={\ensuremath{\underline{\bar{\bs{u}}}_k}},
	plural={\ensuremath{\underline{\bar{\bs{u}}}}},
	description={Nominal/mean vector of the stacked input vector}
}
\newglossaryentry{nomStates}{type=symbols,
	sort={x4},
	name={\ensuremath{\underline{\bar{\bs{x}}}_k}},
	plural={\ensuremath{\underline{\bar{\bs{x}}}}},
	description={Nominal/mean vector of the stacked state vector}
}

\newglossaryentry{sysMat}{type=symbols,
	sort={A},
	name={\ensuremath{\bs{A}}},
	description={System matrix with the dimension $\gls{dimx}\times\gls{dimx}$}
}
\newglossaryentry{inputMat}{type=symbols,
	sort={B},
	name={\ensuremath{\bs{B}}},
	description={Input matrix with the dimension $\gls{dimx}\times\gls{dimu}$}
}
\newglossaryentry{distMat}{type=symbols,
	sort={G},
	name={\ensuremath{\bs{G}}},
	description={Disturbance matrix with the dimension $\gls{dimx}\times\gls{dimw}$}
}
\newglossaryentry{fbMat}{type=symbols,
	sort={K},
	name={\ensuremath{\bs{K}}},
	description={Feedback matrix for \acs{RMPC}}
}

\newglossaryentry{liftedSysMat}{type=symbols,
	sort={A},
	name={\ensuremath{\bs{\underline{A}}_k}},
	plural={\ensuremath{\bs{\underline{A}}}},
	description={Lifted version of the system matrix with the dimension $\gls{dimx}\cdot i\times\gls{dimx}$}
}
\newglossaryentry{liftedInputMat}{type=symbols,
	sort={B},
	name={\ensuremath{\bs{\underline{B}}_k}},
	plural={\ensuremath{\bs{\underline{B}}}},
	description={Lifted version of the input matrix with the dimension $\gls{dimx}\cdot i\times\gls{dimu}\cdot i$}
}
\newglossaryentry{liftedDistMat}{type=symbols,
	sort={G},
	name={\ensuremath{\bs{\underline{G}}_k}},
	plural={\ensuremath{\bs{\underline{G}}}},
	description={Lifted version of the disturbance matrix with the dimension $\gls{dimx}\cdot i\times\gls{dimw}\cdot i$}
}

\newglossaryentry{dimx}{type=symbols,
	sort={nx},
	name={\ensuremath{n_x}},
	description={Dimension of the state vector \gls{state}}
}
\newglossaryentry{dimu}{type=symbols,
	sort={nu},
	name={\ensuremath{n_u}},
	description={Dimension of the input vector \gls{input}}
}
\newglossaryentry{dimw}{type=symbols,
	sort={nw},
	name={\ensuremath{n_x}},
	description={Dimension of the disturbance vector \gls{dist}}
}
\newglossaryentry{Nmpc}{type=symbols,
	sort={n},
	name={\ensuremath{N}},
	description={Horizon for \ac{MPC}}
}
\newglossaryentry{Ncvpm}{type=symbols,
	sort={ncvpm},
	name={\ensuremath{N}},
	description={Horizon for \ac{CVPM}}
}

\newglossaryentry{inputSet}{type=symbols,
	sort={Us},
	name={\ensuremath{\mathcal{U}}},
	description={Set for the input vector \gls{input}}
}
\newglossaryentry{stateSet}{type=symbols,
	sort={Xs},
	name={\ensuremath{\mathcal{X}}},
	description={Set for the state vector \gls{state}}
}
\newglossaryentry{distSet}{type=symbols,
	sort={Ws},
	name={\ensuremath{\mathcal{W}}},
	description={Set for the disturbance vector \gls{dist}}
}
\newglossaryentry{distSetconst}{type=symbols,
	sort={Ws},
	name={\ensuremath{\mathcal{W}}},
	description={Set for the disturbance vector \gls{dist}}
}
\newglossaryentry{terminalSet}{type=symbols,
	sort={Xsf},
	name={\ensuremath{\mathcal{X}_\textnormal{f}}},
	description={Terminal set for the last state vector $\glspl{state}_{\gls{Nmpc}}$}
}

\newglossaryentry{stateSetPolytope}{type=symbols,
	sort={e},
	name={\bs{E}, \bs{e}},
	description={Matrix and vector for the polyhedral representation of state sets}
}

\newglossaryentry{inputSetPolytope}{type=symbols,
	sort={f},
	name={\bs{F}, \bs{f}},
	description={Matrix and vector for the polyhedral representation of input sets}
}

\newglossaryentry{distSetPolytope}{type=symbols,
	sort={l},
	name={\bs{L}, \bs{l}},
	description={Matrix and vector for the polyhedral representation of disturbance sets}
}

\newglossaryentry{adInputSet}{type=symbols,
	sort={Usx},
	name={\ensuremath{\underline{\mathcal{U}}}},
	description={Admissible input set for \acs{MPC}-optimization}
}
\newglossaryentry{adInputSetNp}{type=symbols,
	sort={Usx},
	name={\ensuremath{\underline{\mathcal{U}}}},
	description={Admissible input set for \acs{MPC}-optimization}
}

\newglossaryentry{probStateSet}{type=symbols,
	sort={Xsp},
	name={\ensuremath{\mathcal{X}}},
	description={Constraint for the CVPM optimization. The probability of not being in the set is minimized}
}
\newglossaryentry{probStatesSet}{type=symbols,
	sort={Xsp},
		name={\ensuremath{\underline{\mathcal{X}}}},
	description={Constraint for the CVPM optimization. The probability of not being in the set is minimized}
}
\newglossaryentry{probStateSet_k}{type=symbols,
	sort={Xsp},
	name={\ensuremath{\mathcal{X}_k}},
	description={Constraint for the CVPM optimization. The probability of not being in the set is minimized}
}
\newglossaryentry{probStateSetconst}{type=symbols,
	sort={Xsp},
	name={\ensuremath{\mathcal{X}}},
	description={Constraint for the CVPM optimization. The probability of not being in the set is minimized}
}
\newglossaryentry{probInputSet}{type=symbols,
	sort={Usp},
	name={\ensuremath{\mathcal{U}_{\mathcal{X}}}},
	description={Input set, where the states are in the probabilistic state set \gls{probStateSet}}
}
\newglossaryentry{probInputSetNp}{type=symbols,
	sort={Usp},
	name={\ensuremath{\underline{\mathcal{U}}_{\mathcal{X}}}},
	description={Input set, where the states are in the probabilistic state set \gls{probStateSet}}
}

\newglossaryentry{EqStateSet}{type=symbols,
	sort={Xseq},
	name={\ensuremath{\mathcal{X}_\textnormal{eq}}},
	description={Set of states for which an input exists such that this state is a steady-state}
}
\newglossaryentry{EqInputSet}{type=symbols,
	sort={Useq},
	name={\ensuremath{\mathcal{U}_\textnormal{eq}}},
	description={Set of inputs that leads to a steady-state}
}
\newglossaryentry{distInvSet}{type=symbols,
	sort={Zs},
	name={\ensuremath{\mathcal{Z}}},
	description={Disturbance invariant set}
}
\newglossaryentry{optInputSet}{type=symbols,
	sort={Usopt},
	name={\ensuremath{\underline{\mathcal{U}}_\textnormal{cvpm}}}, 
	description={Input set generated by CVPM}
}
\newglossaryentry{optInputSetNp}{type=symbols,
	sort={Usopt},
	name={\ensuremath{\underline{\mathcal{U}}_{\textnormal{cvpm}}}},
	description={Input set generated by CVPM}
}
\newglossaryentry{targetSet}{type=symbols,
	sort={Xst},
	name={\ensuremath{\mathcal{X}_\textnormal{lp}}},
	description={Target set for the probability optimization}
}
\newglossaryentry{neigTargetSet}{type=symbols,
	sort={Xnst},
	name={\ensuremath{\mathcal{X}_\textnormal{Nlp}}},
	description={Neighborhood of  the  target set \gls{targetSet}}
}
\newglossaryentry{robustSet}{type=symbols,
	sort={Xsr},
	name={\ensuremath{\mathcal{X}_\textnormal{R}}},
	description={Robust set of the probability optimization}
}

\newglossaryentry{case1Set}{type=symbols,
	sort={Xs},
	name={\ensuremath{\mathcal{X}_\textnormal{caseS}}},
	description={Set of initial states for case 1}
}

\newglossaryentry{int_xi}{type=symbols,
	sort={xi},
	name={\ensuremath{\underline{\bs{\xi}}}},
	description={}
}

\newglossaryentry{Sigmaw}{type=symbols,
	sort={sigmaw},
	name={\ensuremath{\bs{\Sigma}_{\glspl{dist}}}},
	description={Covariance matrix of the disturbance vector \gls{dist}}
}

\newglossaryentry{Sigmax}{type=symbols,
	sort={sigmax1},
	name={\ensuremath{\bs{\Sigma}_{\glspl{state}}}},
	description={Covariance matrix of the state $\glspl{state}$}
}

\newglossaryentry{SigmaX}{type=symbols,
	sort={sigmax},
	name={\ensuremath{\bs{\Sigma}_{\glspl{states}}}},
	plural={\ensuremath{\tilde{\bs{\Sigma}}_{\glspl{states}}}},
	description={Covariance matrix of the stacked state vector $\gls{states}$}
}

\newglossaryentry{SigmaW}{type=symbols,
	sort={sigmaw},
	name={\ensuremath{\bs{\Sigma}_{\glspl{dists}}}},
	plural={\ensuremath{\bs{\Sigma}}},
	description={Covariance matrix of the stacked disturbance vector $\gls{dists}$}
}

\newglossaryentry{inputCost}{type=symbols,
	sort={R},
	name={\ensuremath{\bs{R}}},
	description={Weight matrix for inputs within the optimization horizon}
}
\newglossaryentry{stateCost}{type=symbols,
	sort={Q},
	name={\ensuremath{\bs{Q}}},
	description={Weight matrix for states within the optimization horizon}
}
\newglossaryentry{terminalCost}{type=symbols,
	sort={Qt},
	name={\ensuremath{\bs{Q}_\textnormal{f}}},
	description={Weight matrix for last state in the otpimization, i. e. terminal cost}
}
\newglossaryentry{cost}{type=symbols,
	sort={J},
	name={\ensuremath{J}},
	description={Cost function}
}

\newglossaryentry{ominus}{type=notation,
	sort={pontriangin},
	name={\ensuremath{\ominus}},
	description={Pontryagin difference}
}	

\newglossaryentry{oplus}{type=notation,
	sort={Minkowski},
	name={\ensuremath{\oplus}},
	description={Minkowski sum}
}	

\newglossaryentry{probability}{type=notation,
	sort={probability},
	name={\ensuremath{\operatorname{Pr}(\bullet)}},
	description={Probability of a event}
}	

\newglossaryentry{expection}{type=notation,
	sort={expection},
	name={\ensuremath{\operatorname{E}(\bullet)}},
	description={Expection value}
}	

\newglossaryentry{variance}{type=notation,
	sort={variance},
	name={\ensuremath{\operatorname{Var}(\bullet)}},
	description={Variance or Covariance matrix}
}

\def\BibTeX{{\rm B\kern-.05em{\sc i\kern-.025em b}\kern-.08em
    T\kern-.1667em\lower.7ex\hbox{E}\kern-.125emX}}
\begin{document}
\title{
Minimal Constraint Violation Probability in Model Predictive Control for Linear Systems}
\author{Michael Fink, Tim Br\"udigam, Dirk Wollherr, and Marion Leibold
\thanks{This work was funded by the Deutsche Forschungsgemeinschaft (DFG, German Research Foundation) – project number 490649198. This work was supported by a fellowship within the IFI program of the German Academic Exchange Service (DAAD) and the Bavaria California Technology Center (BaCaTeC) grant 1-[2020-2].}
\thanks{The authors are with the Chair of Automatic Control Engineering, Department of Electrical and Computer Engineering, Technical University
	of Munich, 80333 Munich, Germany (e-mail: \{michael.fink, tim.bruedigam, dw, marion.leibold\}@tum.de).}
}

\maketitle

\begin{abstract}
Handling uncertainty in model predictive control comes with various challenges, especially when considering state constraints under uncertainty. 
Most methods focus on either the conservative approach of robustly accounting for uncertainty or allowing a small probability of constraint violation. 
In this work, we propose a linear model predictive control approach that minimizes the probability that linear state constraints are violated in the presence of additive uncertainty.
This is achieved by first determining a set of inputs that minimize the probability of constraint violation. Then, this resulting set is used to define admissible inputs for the optimal control problem. 
Recursive feasibility is guaranteed and input-to-state stability is proved under assumptions. Numerical results illustrate the benefits of the proposed model predictive control approach.
\end{abstract}


\section{Introduction}
\label{sec:introduction}

\glsunset{cvpm}

Considering uncertainty presents a major challenge in the control of safety-critical systems. Depending on the application, uncertainty ranges from noise and disturbances to model and parameter inaccuracy. 
For example, the control of automated vehicles needs to account for sensor noise, disturbances such as wind, or unknown future behavior of other vehicles that are close enough to be considered for safety.
Ideally, the control of such safety-critical systems realizes minimal risk in the presence of uncertainty.

A prominent method for the control of safety-critical systems is \gls{MPC}, due to its ability to consider input and state constraints to satisfy safety requirements \cite{RawlingsMayneDiehl2017}. 
In general \gls{MPC} requires a model of the system to solve an optimal control problem in each time step.

When uncertainty is present, constraints are handled in a robust way by \gls{RMPC} \cite{BemporadMorari1999}. Initially known bounds on the uncertainty allow for a guarantee on stability and recursive feasibility. 
Nevertheless, robustly accounting for uncertainty comes with issues. If the uncertainty bound was initially not estimated large enough, all guarantees are lost. If uncertainty bounds are chosen too large, potentially to account for rare worst-case events, \gls{RMPC} becomes highly conservative.

Overcoming conservatism is addressed by \gls{SMPC} \cite{Mesbah2016, FarinaGiulioniScattolini2016}. In \gls{SMPC}, constraints subject to uncertainty are handled as chance constraints. A chance constraint requires that the constraint is satisfied to a certain level, based on a chosen risk parameter, representing the acceptable risk. While a low-risk parameter results in a high probability of constraint violation, performance is improved, as rare uncertainty realizations are neglected. Again, multiple issues arise. Similar to \gls{RMPC}, wrong initial assumptions for the uncertainty cause feasibility issues. 
Furthermore, \gls{SMPC} does not penalize the allowed constraint violation, e.g. the constraint violation probability is not considered in the cost function.

When developing MPC for safety-critical systems, a small probability of constraint violation may be tolerated. It is nevertheless fundamental to achieve the minimum probability of constraint violation.
In addition, recursive feasibility and stability of the closed-loop system dynamics are essential. Also dealing with variations in the uncertainty as well as in the constraints has to be considered.

Whereas \gls{RMPC} and \gls{SMPC} partially consider these requirements, both methods are impractical for safety-critical systems due to the above-mentioned issues, e.g., uncertainty bounds must be known in advance or recursive feasibility is not guaranteed in every situation. 
These considerations resulted in the development of \gls{MPC} with constraint violation probability minimization (CVPM) \cite{BruedigamEtalLeibold2021c,Fink2022}. 

\subsection{Related Work}

While there exist different approaches to \gls{RMPC}, the main ideas are similar: robustly handling constraints, accounting for worst-case uncertainty realizations. An overview of \gls{RMPC} approaches is given in \cite[Ch.~3]{RawlingsMayneDiehl2017} and a summary of early works in \cite{MayneEtalScokaert2000}. The most prominent \gls{RMPC} approaches are min-max MPC \cite{RaimondoEtalCamach2009}, considering maximal uncertainties, and tube-based MPC \cite{LangsonEtalMayne2004, KoehlerEtalAllgoewer2021}, which defines a tube around the nominal state trajectory to tighten constraints appropriately. 
While \gls{RMPC} was used to control safety-critical systems, e.g., a robotic manipulator \cite{NubertEtalTrimpe2020} or an automated vehicle \cite{GaoEtalBorrelli2014, SolopertoEtalMueller2019}, solutions are generally conservative, as no knowledge about the disturbance distribution is taken into account. Additionally, changing uncertainty bounds or distributions, as well as changing constraints cause a loss of properties, e.g., recursive feasibility or require intensive recomputation.

\gls{SMPC} employs chance constraints, i.e., the probability of violating a constraint is bounded by a risk parameter.
Chance constraints are difficult to handle in general as their evaluation requires computing multivariate integrals. Therefore, approaches to reformulating these probabilistic constraints into deterministic (potentially approximated) constraints are proposed:
For Gaussian uncertainties and linear constraints, the chance constraints can be analytically reformulated \cite{FarinaGiulioniScattolini2016}. Gaussian distributions are suitable to describe certain types of uncertainty, e.g., noise, while often safety-critical systems require considering other distributions. For this case, sampling-based \gls{SMPC} approaches are suitable. Particle-based \gls{SMPC} \cite{BlackmoreEtalWilliams2010} approximates the chance constraint by ensuring that only a small number of samples leads to constraint violations. In \gls{SCMPC} \cite{SchildbachEtalMorari2014}, based on \cite{CampiGaratti2011}, the number of required samples is computed depending on a risk parameter. Then, for all sampled scenarios, the constraints must be satisfied. A major issue with sampling-based \gls{SMPC} is that guarantees on recursive feasibility are difficult to obtain. 
A combination of \gls{RMPC} and {SMPC} is considered in \cite{BruedigamEtalLeibold2020b, BruedigamEtalBuss2021}, where robust constraints are employed on the short-term horizon and chance constraints are used for long-term predictions. This approach improves performance compared to \gls{RMPC} while still employing short-term robust constraints; however, proofs of recursive feasibility and stability are challenging. 

The \gls{SMPC} approaches, introduced in the previous paragraph, consider open-loop chance constraints, i.e., chance constraint satisfaction is only required in the optimal control problem for the open-loop prediction. Though, recursive feasibility can only be guaranteed if closed-loop constraint satisfaction is considered \cite{HewingWabersichZeilinger2020}. An alternative method to guarantee recursive feasibility is proposed in \cite{LorenzenEtalAllgoewer2017}, proposing an additional constraint on the predicted state of the first prediction step, making recursive feasibility challenging in SMPC.

\gls{SMPC} is mostly applied to applications where constraint violations are not critical, e.g., energy control in buildings \cite{OldewurtelEtalMorari2014, MaMatuskoBorrelli2015} or hybrid electrical vehicles \cite{diCairanoEtalKolmanovsky2014, ZengWang2015}. \gls{SMPC} for safety-critical systems mostly focused on automated vehicles 
\cite{SchildbachBorrelli2015, CesariEtalBorrelli2017, BruedigamEtalLeibold2020c, BruedigamEtalLeibold2021d}. 

However, safety of \gls{SMPC} is only specifically addressed in few works. In \cite{BruedigamEtalLeibold2021b}, failsafe trajectory planning is used to guarantee safety in case of infeasible \gls{SMPC} optimal control problems. A different idea is presented in \cite{ZhangLinigerBorrelli2021}, where a least-intrusive trajectory is found if a collision is inevitable. 

All previously described \gls{SMPC} approaches are unable to provide recursive feasibility or stability guarantees once unexpected changes arise, potentially due to time-varying uncertainties or time-varying constraints.
While slack variables can be introduced or alternative problems can be solved \cite{CannonKouvaritakisWu2009, CannonKouvaritakisWu2009b} to regain feasibility, 
the optimization does not necessarily provide the safest possible solution.  
This issue is addressed in \cite{BruedigamEtalLeibold2021c} for a collision avoidance scenario. There, the probability of violating a collision avoidance norm constraint in the first prediction step is minimized with \gls{cvpm}. 
An extension of CVPM-MPC where the norm constraint is taken into account for more than one prediction step is presented in \cite{Fink2022}.
Unlike multi-objective MPC \cite{WojsznisEtalBlevins2007,BemporadMunoz2009}, in CVPM-MPC, safety is not part of an objective trade-off, but safety is maximized first before other objectives are considered. However, the approach in \cite{BruedigamEtalLeibold2021c} and \cite{Fink2022} is designed particularly for obstacle avoidance, i.e., the approach is limited to norm constraints, and only uncertainty affecting the obstacle dynamics is considered. 

In summary, previous \gls{MPC} approaches only cover parts of the requirements for safety-critical systems. 
The major challenge, reasonably minimizing constraint violation probabilities for general linear systems, is still an open problem.

\subsection{Contribution and Structure}
In this work, we propose an MPC algorithm that minimizes the probability of constraint violation. This is achieved by guaranteeing constraint satisfaction whenever possible and ensuring minimal constraint violation probability whenever constraint satisfaction is not possible. 
We generalize, extend, and simplify the results of the \gls{cvpm} method from \cite{BruedigamEtalLeibold2021c} and \cite{Fink2022}. In particular, we now replace norm constraints by general  linear constraints.

We investigate the requirements for recursive feasibility and stability. 
We show that no additional assumptions are needed for recursive feasibility because there are no bounds on the maximum allowed constraint violation probability.  
Furthermore, we propose additional rather strong assumptions for proving input-to-state stability (ISS).  
Finally, the practical consequences of relaxing these assumption are discussed.

The remaining parts are structured as follows. 
Section~\ref{sec:problem} introduces the problem. The \gls{cvpm} method for linear systems and constraints is derived in Section~\ref{sec:method}, followed by details on the properties in Section~\ref{sec:properties}. A numerical example is shown in Section~\ref{sec:results}, demonstrating the benefits of applying \gls{cvpm} to safety-critical systems. A discussion and conclusive remarks are given in Section~\ref{sec:discussion} and Section~\ref{sec:conclusion}, respectively.

\subsection{Notation}
Norms are denoted by $||.||$. We define $||\bs{a}||_{\bs{A}}^2 = \bs{a}^\top \bs{A} \bs{a}$. An augmented vector is denoted by $\underline{\bs{a}}= [ \bs{a}_1, \cdots, \bs{a}_i ]^\top $. We denote linear transformations of sets by $\bm{A} \circ \mathcal{B} = \setdeff{\bm{A} \bm{b}}{\bm{b} \in \mathcal{B}}$ and $\mathcal{B} \circ \bm{A} = \setdeff{\bm{b}}{\bm{A}\bm{b} \in \mathcal{B}}$. 
The Cartesian product of the set $\mathcal{A}$ and $\mathcal{B}$ is $\mathcal{A}\times\mathcal{B}= \left\{ [\bs{a},\bs{b}] \;\middle|\; \bs{a}\in\mathcal{A},\bs{b}\in\mathcal{B}\right\}$.
The $n$-ary Cartesian power of a set $\mathcal{A}$ is denoted by $\mathcal{A}^n = \setdeff{ [ \bm{a}_1, \cdots, \bm{a}_i ] }{\bm{a}_i \in \mathcal{A}~ \forall i \in \intl{1}{n}}$. The Minkowski sum of two sets is denoted $\mathcal{A} \oplus \mathcal{B} = \left\lbrace \bs{a} + \bs{b} \;\middle|\; \bs{a} \in \mathcal{A}, \bs{b} \in \mathcal{B} \right\rbrace$ and the Pontryagin difference is given by $\mathcal{A} \ominus \mathcal{B} = \left\lbrace \bs{c} \;\middle|\; \bs{c} + \bs{b} \in \mathcal{A}, \forall \bs{b} \in \mathcal{B} \right\rbrace$. 
A state at time step $t$ is denoted by $\bs{x}_t$. Within an \gls{MPC} optimal control problem, the state at prediction step $k$ is denoted by $\bs{x}_{k+t}$, where the notation $\bs{x}_{k+t|t}$ explicitly denotes that the prediction $\bs{x}_{k+t}$ was obtained at time step $t$. 
A function $\gamma: \mathbb{R}^n \rightarrow \mathbb{R}_{\geq0}$ is of class $\mathcal{K}$ if it is strictly increasing and $\gamma(0) = 0$. A function $\alpha: \mathbb{R}^n \rightarrow \mathbb{R}_{\geq0}$ is of class $\mathcal{K}_{\infty}$ if $\alpha \in \mathcal{K}$ and $\lim_{s \rightarrow \infty} \alpha(s) = \infty$. 
An asterisk denotes the optimal value, i.e., $\bm{u}^*$ is the optimal solution.

\section{Problem Setup}
\label{sec:problem}

In the following, we first introduce the system dynamics and define properties of the uncertainty. Then, the \gls{MPC} problem statement is introduced, including constraints for which violation probability must be minimized.

\subsection{System Dynamics}

We consider the linear, discrete-time dynamical control system
\begin{IEEEeqnarray}{c}
\glspl{state}_{t+1} = \gls{sysMat}\gls{state}  + \gls{inputMat}\gls{input} + \gls{distMat}\gls{dist} \label{eq:sys}
\end{IEEEeqnarray}
with state $\gls{state}\in \mathbb{R}^{\gls{dimx}}$ and input $\gls{input}\in \mathbb{R}^{\gls{dimu}}$ at time step $t$, as well as the bounded uncertainty $\gls{dist} \in\gls{distSet}\subseteq\mathbb{R}^{\gls{dimw}}$, where $\gls{sysMat},  \gls{inputMat}$ have appropriate dimensions, $\gls{distMat}\in\mathbb{R}^{\gls{dimx}\times\gls{dimx}}$ is not singular. 

\begin{assumption} \label{ass:uncertainy}
The uncertainty \gls{dist} is a truncated Gaussian uncertainty with $\gls{dist} ~\sim \mathcal{N}\lr{\bs{0},\gls{Sigmaw}}$, covariance matrix \gls{Sigmaw}, and bounded by the polytope  \gls{distSet}.
\end{assumption}

In the following, MPC requires to predict the state trajectory. Based on the initial state $\glspl{state}_t$, we denote the augmented system dynamics, delivering predictions $\glspl{state}_{t+1}, \glspl{state}_{t+2}, ... ,\glspl{state}_{t+N}$, by
\begin{IEEEeqnarray}{c}
\glspl{states}  = \glspl{liftedSysMat}\,\glspl{state}_t  + \glspl{liftedInputMat}\,\glspl{inputs} + \glspl{liftedDistMat}\,\glspl{dists} \label{eq:sys_lifted}
\end{IEEEeqnarray}
\begin{IEEEeqnarray}{c}
\text{where }
\glspl{states}=\begin{bmatrix} \glspl{state}_{t+1} \\: \\ \glspl{state}_{t+\gls{Nmpc}} \end{bmatrix}, \quad
\glspl{inputs}=\begin{bmatrix} \glspl{input}_{t} \\ : \\ \glspl{input}_{t+\gls{Nmpc}-1} \end{bmatrix}, \quad
\glspl{dists}=\begin{bmatrix} \glspl{dist}_t \\ : \\ \glspl{dist}_{t+\gls{Nmpc}-1} \end{bmatrix} \IEEEeqnarraynumspace 
\end{IEEEeqnarray}
and \glspl{liftedSysMat}, \glspl{liftedInputMat}, and \glspl{liftedDistMat} defined as in \cite{BorrelliBemporadMorari2017}.

\subsection{Model Predictive Control}

\gls{MPC} is applied to control system~\eqref{eq:sys}. In \gls{MPC}, an optimal control problem is repeatedly solved on a finite horizon $N$, where state and input constraints are considered and only the first entry of the optimal input trajectory is applied. 
The prediction steps are denoted by index $k$. The \gls{MPC} cost function is given by
\begin{IEEEeqnarray}{c}
\gls{cost}(\glspl{state}_t,\glspl{inputs}) = \sum_{k=0}^{N-1}\lrsum{ ||\overline{\glspl{state}}_{t+k}||_{\gls{stateCost}}^2 + ||\overline{\glspl{input}}_{t+k}||_{\gls{inputCost}}^2 } +||\overline{\glspl{state}}_{t+N}||_{\gls{terminalCost}}^2 \label{eq:cost}
\end{IEEEeqnarray}
with weighting matrices \gls{stateCost}, \gls{inputCost} and terminal weighting matrix \gls{terminalCost}, where $\glspl{state}_t$ is known and $\overline{\glspl{state}}_{t+k}$ denotes the mean of $\glspl{state}_{t+k}$.

Furthermore, the set of admissible input sequences  \glspl{inputs} is given as the  bounded polytopic set  $\gls{inputSet}^{\gls{Nmpc}}$, which contains the origin.
The polytopic set ${\gls{probStateSet}\subset\mathbb{R}^{\gls{dimx}}}$ allows to formalize state constraints for all states in a prediction horizon of length \gls{Ncvpm}.

\begin{assumption}\label{ass:xP_origin}
The constraint set $\gls{probStateSet}$ is closed, bounded, and contains the origin.
\end{assumption}
The constraint set \gls{probStateSet} may be expressed in augmented form by
\begin{IEEEeqnarray}{c} \label{eq:constraintNp}
\glspl{states} \in\gls{probStatesSet} 
= \gls{probStateSet}^{\gls{Ncvpm}-1}\times \gls{terminalSet}.
\end{IEEEeqnarray}
A terminal constraint $\glspl{state}_N\in\gls{terminalSet}\subseteq\gls{probStateSet}$ is introduced  for the state $\glspl{state}_N$. The set \gls{terminalSet} is defined later as part of the stability analysis.

\subsection{Problem Statement}

Both, \gls{RMPC} and \gls{SMPC}, do not provide adequate solutions if the probability of violating constraint \eqref{eq:constraintNp} must be minimized. This problem may be formulated as
\begin{IEEEeqnarray}{c}
\min_{\glspl{inputs}\in\gls{inputSet}^{\gls{Ncvpm}}}\operatorname{Pr}\left(\glspl{states}\notin \gls{probStatesSet}\right).
\end{IEEEeqnarray}
Given this constraint violation probability minimization, we now specify the problem to be addressed in this work.

\begin{problem}\label{prob:cvpm}
The optimal control problem of each MPC iteration corresponding to time $t$ is
\begin{IEEEeqnarray}{rl} \IEEEyesnumber\label{eq:MPCunderlying}
&\min_{ \glspl{inputs} \in \gls{inputSet}^{\gls{Ncvpm}}}\gls{cost}(\glspl{state}_t,\glspl{inputs}) \IEEEyessubnumber  \\
\text{s.t}&\quad \glspl{inputs} = \arg\min_{\glspl{inputs}\in\gls{inputSet}^{\gls{Ncvpm}}}\operatorname{Pr}\left(\glspl{states}\notin\gls{probStatesSet}\right) .
\IEEEyessubnumber \label{eq:proboptim0} 
\end{IEEEeqnarray}
\end{problem}

In the following section, an \gls{MPC} method is derived that provides a strategy to solve Problem~\ref{prob:cvpm}.
\section{Method}
\label{sec:method}

Here, we first present the CVPM-MPC method in Section~\ref{sec:cvpm_mpc_method} and then details on the probability optimization required for CVPM-MPC in Section~\ref{sec:proboptim}. Major properties of the \gls{cvpm} method are discussed in Section~\ref{sec:properties}. 

\subsection{CVPM-MPC}
\label{sec:cvpm_mpc_method}

The general idea of \gls{cvpm} is to solve an \gls{MPC} optimal control problem where only those inputs are allowed that enable minimal constraint violation probability. 
Therefore, a set is defined that includes all inputs that minimize the constraint violation probability.

\begin{definition}[Optimal CVPM Input Set] \label{def:cvpm}
	The optimal CVPM input set \gls{optInputSet} consists of admissible input sequences $\glspl{inputs}$ that minimize the constraint violation probability $\operatorname{Pr}\left(\glspl{states}\notin\gls{probStatesSet}\right)$. 
\end{definition}

Determining \gls{optInputSet} uses two cases, depending on whether an input sequence $\glspl{inputs}$ exists that guarantees constraint satisfaction or not.

\begin{definition}[CVPM \gls{CC1}] \label{def:c1}
	In \gls{c1}, at least one admissible input sequence $\glspl{inputs}$ exists that guarantees constraint satisfaction, i.e.,
	\begin{IEEEeqnarray}{c}
		\exists\; \glspl{inputs} \in \gls{inputSet}^{\gls{Ncvpm}} \text{ s.t. } \operatorname{Pr}\left(\glspl{states}\notin\gls{probStatesSet}\right) = 0 .
	\end{IEEEeqnarray}
\end{definition}
\begin{definition}[CVPM \gls{CC2}] \label{def:c2}
	In \gls{c2}, no input sequence $\glspl{inputs}$ exists that guarantees constraint satisfaction, i.e., 
	\begin{IEEEeqnarray}{c}
		\operatorname{Pr}\left(\glspl{states}\notin\gls{probStatesSet}\right) > 0 \; \forall \; \glspl{inputs} \in \gls{inputSet}^{\gls{Ncvpm}} .
	\end{IEEEeqnarray}
\end{definition}

At each MPC iteration, it is determined whether the safe case is feasible, i.e. it is evaluated if the set of possible input sequences is not empty. If not, the probabilistic case is applied. 
In the following, for both cases separately, it is addressed how the set \gls{optInputSetNp} is obtained.

\subsubsection{\gls{CC1}}
Here the set \gls{optInputSetNp} is 
\begin{IEEEeqnarray}{c}
	\gls{optInputSet}=\gls{inputSet}^{\gls{Ncvpm}} \cap  \left(
	\gls{probStatesSet}\ominus \lr{\glspl{liftedDistMat}\circ\gls{distSet}^{\gls{Ncvpm}}} \oplus \left\lbrace-\glspl{liftedSysMat}\glspl{state}_t\right\rbrace
	\right) \circ \glspl{liftedInputMat}  ,\label{eq:Uopt_cases}
\end{IEEEeqnarray}
which is the intersection of the admissible input set $\gls{inputSet}^{\gls{Ncvpm}}$ and the set of input sequences that guarantee constraint satisfaction of \eqref{eq:constraintNp}. 
The set $\left\lbrace-\glspl{liftedSysMat}\glspl{state}_t\right\rbrace$ is a singleton that takes into account the current state $\gls{state}$ and the set \gls{optInputSet} is determined with algorithms from \cite{BorrelliBemporadMorari2017}. 

The intersection \eqref{eq:Uopt_cases} does not only deliver the set of admissible inputs, it also allows to check if the safe case applies: if \gls{optInputSetNp} is non-empty, an input sequence \glspl{inputs} exists guaranteeing that the constraint \eqref{eq:constraintNp} is satisfied.

Based on the previous result, it is possible to obtain a set of feasible initial states \gls{case1Set} (for the current MPC iteration) for which \gls{c1} is applicable. This set is given by
\begin{equation}
\gls{case1Set} = \left( \gls{probStatesSet} \ominus \lr{\glspl{liftedDistMat}\circ \gls{distSet}^{\gls{Ncvpm}}} \oplus \lr{\left(-\glspl{liftedInputMat} \right)\circ\gls{inputSet}^{\gls{Ncvpm}}} \right)\circ \glspl{liftedSysMat}, \label{eq:case1Init}
\end{equation}
which is obtained analog to \eqref{eq:Uopt_cases}.
We make the following assumption to ensure that \gls{case1Set} is non-empty.
\begin{assumption}\label{ass:distSubX}
    The disturbances are small enough and propagated disturbances alone never exceed state constraints, i.e.,
    $\glspl{liftedDistMat}\circ \gls{distSet}^{\gls{Ncvpm}} \subset\gls{probStatesSet}.$
\end{assumption}
This assumption is straight-forward to satisfy when the system matrix \gls{sysMat} is stable and motivates the following assumption. 
\begin{assumption}\label{ass:stabelA}
	The system matrix \gls{sysMat} is stable, i.e. the eigenvalues of \gls{sysMat} are within the unit circle.
\end{assumption}
\begin{remark}\label{rem:prestab}
Note that even if the system is unstable, the assumption on \gls{sysMat} can still be fulfilled by using a feedback controller for prestabilization.
Then the sets \gls{stateSet} and \gls{inputSet} have to be redefined taking into account the prestabilization, such as in \cite{Xu2023}.
\end{remark}

\subsubsection{\gls{CC2}}
For \gls{c1}, \gls{optInputSetNp} collects those input trajectories that guarantee constraint satisfaction. However, if such an input trajectory does not exist, at least minimal constraint violation probability can be ensured. Therefore, for \gls{c2}, \gls{optInputSetNp} collects input trajectories $\glspl{inputs}^*$ that result in minimal constraint violation probability.

\gls{C1} is applied if the current state is in \gls{case1Set} and results in zero constraint violation probability. 
Thus for all other states, i.e., $\gls{state}\notin \gls{case1Set}$, the probability of constraint violation is non-zero. 
To minimize the probability of violating $\glspl{states}\in  \gls{probStateSet}^{\gls{Ncvpm}}$, we transform the problem to minimize the probability of violating  $\glspl{states}\in  \gls{case1Set}^{\gls{Ncvpm}}$ 
\begin{IEEEeqnarray}{rl}
	\IEEEyesnumber \label{eq:proboptim}
	\glspl{inputs}^* =&\arg\min_{\glspl{inputs}\in  \gls{inputSet}^{\gls{Ncvpm}} }\operatorname{Pr}\left( \glspl{states} \notin\gls{case1Set}^{\gls{Ncvpm}} \right) .
\end{IEEEeqnarray}

The set ${\gls{stateSet}\setminus\gls{case1Set}}$ contains all states in \gls{stateSet} that are not initial states for \gls{c1}, where all predicted states remains in \gls{probStatesSet}. 
A trajectory will leave \gls{probStatesSet} eventually if its initial state is in ${\gls{stateSet}\setminus\gls{case1Set}}$. 
Therefore, trajectories with states in ${\gls{stateSet}\setminus\gls{case1Set}}$ result in a high constraint violation probability. 
In contrast, ${\gls{case1Set}\setminus\gls{stateSet}}$ contains all states that are not in \gls{stateSet}, but the predicted trajectory remains in \gls{probStatesSet} and results in a zero probability of constraint violation in the subsequent MPC iterations. 
Changing the probability optimization from \eqref{eq:proboptim0} to \eqref{eq:proboptim} does, therefore, not significantly change  $\glspl{inputs}^*$. However, this adjustment makes it possible to prove stability (see Section~\ref{sec:stability}). 
Once $\glspl{inputs}^*$ is obtained, we set: 
\begin{IEEEeqnarray}{c}
	\gls{optInputSetNp}=\{\glspl{inputs}^* \}. \label{eq:UcvpmCase2}
\end{IEEEeqnarray}

\subsubsection{CVPM-MPC Formulation}
The \gls{cvpm} optimal control problem is 
\begin{IEEEeqnarray}{rl}
	\IEEEyesnumber\label{eq:cvpmoptim}
	&\glspl{inputs}^*=\arg\min_{\glspl{inputs} \in  \gls{optInputSet}}\gls{cost}(\glspl{state}_t,\glspl{inputs})
\end{IEEEeqnarray} 
with \gls{optInputSet} according to \eqref{eq:Uopt_cases} or \eqref{eq:UcvpmCase2}.
The closed-loop system, compare \eqref{eq:sys}, is then given by
\begin{IEEEeqnarray}{c}
	\glspl{state}_{t+1} = \gls{sysMat}\gls{state}  + \gls{inputMat}\bm{u}_t^* + \gls{distMat}\gls{dist} \label{eq:sys_cl}
\end{IEEEeqnarray}
where $\bm{u}_t^*$ is the first element of the optimal trajectory $\glspl{inputs}^*$ obtained at time step~$t$. 

\begin{remark}
Solving  \eqref{eq:cvpmoptim} requires two steps: first, a set  \gls{optInputSet}  is calculated, and then, this set is used as a constraint for the MPC controller. 
Thus, CVPM can also serve as preprocessing for other controllers.
\end{remark}

\subsection{Probability Optimization in \gls{tCC2}}
\label{sec:proboptim}

In \gls{c2}, the input trajectory $\glspl{inputs}^*$ with minimal constraint violation probability, is solution of the optimization problem \eqref{eq:proboptim}. 
In general, neither analytic solution nor exact numerical solution of  $\operatorname{Pr}\left(\glspl{states}\notin\gls{case1Set}^{\gls{Ncvpm}}\right)$ is possible.
Therefore, we propose two approximations.
In the first approximation, the probabilities are computed using a sampling-based approach, which allows to increase the accuracy as the number of samples increases. However, this computation is time-consuming and not suitable for fast real-time systems. Therefore, the second method does not approximate the probability but modifies the optimization problem to find the input sequence \glspl{inputs}.

In the following, we prepare both approximations.
We assume, the probability distribution in \gls{c1} has a truncated support. 
This ensures existence of trajectories with a zero probability of constraint violation.
However, in \gls{c2}, this leads to a vanishing gradient in the optimization. Therefore, we approximate the truncated Gaussian of Assumption~\ref{ass:uncertainy} by the corresponding non-truncated Gaussian ${\gls{dist}\sim \mathcal{N}\lr{\bs{0},\gls{Sigmaw}}}$. The mean $\glspl{nomStates}$ of the state trajectory $\glspl{states}$ is
$ {\glspl{nomStates} =\glspl{liftedSysMat}\gls{state} +\glspl{liftedInputMat}\,\glspl{inputs}} $
and the covariance matrix is given as 
$ {\gls{SigmaX} =\mathrm{diag}\lr{\gls{Sigmax},.., \gls{Sigmax}},}$
where \gls{Sigmax} is the steady-state solution of the uncertainty propagation
\begin{IEEEeqnarray}{c}
	\gls{Sigmax} = \gls{sysMat}\gls{Sigmax}\gls{sysMat}^\top + \gls{distMat}\gls{Sigmaw}\gls{distMat}^\top.
\end{IEEEeqnarray}
Given the mean and the covariance matrix for the state sequence $\glspl{states}$ with a particular input sequence \glspl{inputs}, we obtain
\begin{IEEEeqnarray}{c}
	\glspl{states} \sim \mathcal{N}\left(\glspl{nomStates},\gls{SigmaX}\right)
	= \mathcal{N}\left(\glspl{liftedSysMat}\gls{state} +\glspl{liftedInputMat}\,\glspl{inputs},\gls{SigmaX}\right)
	, \label{eq:statesequenceGauss}
\end{IEEEeqnarray}
proving that the state trajectory is subject to a Gaussian distribution.

\subsubsection{Sampling-Based Probability Optimization} \label{sec:numProb}
A numerical Monte Carlo sampling approach is employed to determine the probability of violating constraints. 
From the distribution \eqref{eq:statesequenceGauss}, $N_\text{s}$ samples of state sequences are drawn. 
The number of samples that are an element of the set \gls{probStatesSet} is denoted as $N_{\gls{probStatesSet}}$. 
It follows that the constraint violation probability is approximately 
\begin{align}\label{eq:montecarlo}
	\Pr\left(\glspl{states}\notin \gls{probStatesSet}\right) \approx 1-\frac{N_{\gls{probStatesSet}} }{N_\text{s}}.
\end{align}
The minimization of the constraint violation probability is approximated with a numeric optimization of  \eqref{eq:montecarlo}.
In each step within the optimization, \eqref{eq:montecarlo} must be determined for a given input sequence \glspl{inputs} resulting in a huge computational burden. 

\subsubsection{Probability Optimization approximated as Quadratic Program} \label{sec:approx}
We propose an approximation in the following to speed up the computation.
For \gls{c2}, the input sequence $\glspl{inputs}^*$ is defined as the solution to the optimization problem~\eqref{eq:proboptim} with 
\begin{IEEEeqnarray}{c}
	\IEEEyesnumber \label{eq:probXinX}
	\operatorname{Pr} \left(\glspl{states}\notin\gls{case1Set}^{\gls{Ncvpm}}\right)  =1 - c
	\int\limits_{ \gls{case1Set}^{\gls{Ncvpm}} } 
	e^{\left(-\frac{1}{2} \left( \glspl{nomStates} - \gls{int_xi} \right)^\top \gls{SigmaX}^{-1} \left( \glspl{nomStates} - \gls{int_xi} \right) \right)} 
	\text{d}\gls{int_xi} ,  \IEEEeqnarraynumspace \IEEEyesnumber
\end{IEEEeqnarray}
where $	c =\lr{(2\pi)^{\gls{dimx}\gls{Ncvpm}} \det \gls{SigmaX}}^{-1}$.
As \eqref{eq:probXinX} is a non-convex cost for the optimization problem, we approximate it. 
In \gls{c2}, the mean state sequence $\glspl{nomStates}$ is not in $\gls{case1Set}^{\gls{Ncvpm}}$, otherwise \gls{c1} is applied.
Therefore, we assume that the distribution function \eqref{eq:statesequenceGauss} is nearly constant over the set $\gls{case1Set}$.
We then approximate the integral in \eqref{eq:probXinX} by a multiplication of the probability density function for the states in \eqref{eq:statesequenceGauss} and the volume of the polytope $\operatorname{V}_\text{P}(\gls{case1Set}^{\gls{Ncvpm}})$ , yielding 
\begin{IEEEeqnarray}{rl}
	\IEEEyesnumber \label{eq:probXinX_approx}
	\operatorname{Pr}&\left(\glspl{states}\notin\gls{case1Set}^{\gls{Ncvpm}}\right)  \approx 1 \! - \! c \,
	e^{\left(-\frac{1}{2} \left( \glspl{nomStates} - \gls{int_xi} \right)^\top \gls{SigmaX}^{-1} \left( \glspl{nomStates} - \gls{int_xi} \right) \right)}
	\!
	\operatorname{V}_\text{P}(\gls{case1Set}^{\gls{Ncvpm}}).    \IEEEeqnarraynumspace 
\end{IEEEeqnarray}
The probability density function is evaluated at a point ${\gls{int_xi} \in \gls{case1Set}^{\gls{Ncvpm}}}$ within the polytope.
The variable \gls{int_xi} is then included in the optimization, yielding a similar structure as the MPC cost function. Later, this structure is essential for the proof of stability. 
The volume of the polytope $\operatorname{V}_\text{P}(\gls{case1Set})$ does not change if \glspl{inputs} is varied.
We need to minimize  $\operatorname{Pr}\left(\glspl{states}\notin\gls{case1Set}^{\gls{Ncvpm}}\right)$, thus it is sufficient to solve the quadratic optimization problem
\begin{IEEEeqnarray}{rl}
	\IEEEyesnumber \label{eq:probOptimMalahanobis}
	\glspl{inputs}^* =&\arg\min_{\glspl{inputs}\in \gls{inputSet}^{\gls{Ncvpm}}\!,\,\gls{int_xi}\in \gls{case1Set}^{\gls{Ncvpm}}}\left( \glspl{nomStates} - \gls{int_xi} \right)^\top \glspl{SigmaX}^{-1} \left( \glspl{nomStates} - \gls{int_xi} \right) \IEEEyessubnumber\\
	\text{s.t. }\quad&  
	\glspl{nomStates} = \glspl{liftedSysMat}\glspl{state}_t+\glspl{liftedInputMat}\glspl{inputs} . \IEEEyessubnumber
\end{IEEEeqnarray}
The matrix \glspl{SigmaX} is an adapted version of \gls{SigmaX} and is defined later when stability is discussed.  
The solution of \eqref{eq:probOptimMalahanobis} leads to state sequences in the direction of small eigenvalues of $ \glspl{SigmaX}^{-1}$, which result in the fastest decrease of the probability of constraint violation.
\section{Properties}
\label{sec:properties}

\subsection{Recursive Feasibility}\label{sec:rf}
Recursive feasibility is a fundamental property for \gls{MPC} algorithms. In each MPC time step, an optimal control problem has to be solved, and it needs to be ensured that the optimal control problem is feasible at time step $t+1$ if it is feasible at time step $t$. 
\begin{definition}[Recursive Feasibility]\label{def:recfeas}
	An \gls{MPC} optimal control problem is recursively feasible if it holds that
	$ {\gls{adInputSet}_t \neq \emptyset \rightarrow \gls{adInputSet}_{t+1} \neq \emptyset}$ 
	for all $t \in \mathbb{N}_{0}$ where $\gls{adInputSet}_t$ is the set of admissible inputs at time step~$t$.
\end{definition}

Recursive feasibility is fulfilled in CVPM per construction, because if \gls{c1} is infeasible, \gls{c2} is applied that will always deliver a solution.
In the following, we prove recursive feasibility of \gls{c1}, i.e., once \gls{c1} is applicable,  \gls{case1Set} is invariant under CVPM-MPC.
For this purpose, the following assumption is required, which defines the terminal set in \eqref{eq:constraintNp}.
\begin{assumption} \label{ass:rci}
	The terminal set $\gls{terminalSet} \subseteq \gls{probStateSet}$ is chosen as a robust control invariant set \cite{BorrelliBemporadMorari2017}	and fulfills 
	\begin{align}\label{eq:rci}
	\gls{sysMat}\circ\gls{terminalSet} \oplus \gls{sysMat}^{\gls{Ncvpm}}\gls{distMat}\circ\gls{distSet} \subseteq \gls{terminalSet} \oplus (-\gls{inputMat})\circ \gls{inputSet}.
	\end{align}
\end{assumption}
\begin{remark}
	Ass.~\ref{ass:rci} yields candidates for the terminal set \gls{terminalSet} by extending the standard notion for a robust invariant set $\gls{sysMat}\circ\gls{terminalSet}\subseteq \gls{terminalSet}  \ominus \gls{distMat}\circ\gls{distSet} \oplus (-\gls{inputMat})\circ \gls{inputSet}$ \cite{BorrelliBemporadMorari2017}, such that disturbances propagated over the horizon \gls{Ncvpm} are included, i.e. it is a robust control invariant set where the set of disturbances is $\gls{sysMat}^{\gls{Ncvpm}}\gls{distMat}\circ\gls{distSet}$. 
\end{remark}

\begin{lemma}\label{lem:rci}
    If Assumption \ref{ass:rci} holds, then for all $\gls{state}\in$ \gls{case1Set} there exists an input \gls{input} such that the state at the next time step is also in the set \gls{case1Set}, i.e., \gls{case1Set} is  robust control invariant.
\end{lemma}

\begin{proof}
	The proof is given in Appendix~\ref{sec:appendic}.
\end{proof}

Based on Lemma~\ref{lem:rci}, we can now formulate the following theorem on recursive feasibility. 

\begin{theorem}\label{th:CVPM}
	\gls{C1} is recursively feasible.
\end{theorem}

\begin{remark}
	In a practical situation, a terminal set, not satisfying Assumption~\ref{ass:rci} is also possible,
	This would imply that \gls{case1Set} is not robust control invariant and thus the recursive feasibility guarantee in \gls{c1} is lost.
	Nevertheless, still a solution with possibly non zero probability of constraint violation can be found. 
\end{remark}

\subsection{Stability}\label{sec:stability}

We show that both \gls{c1} and \gls{c2} ensure input-to-state stability, 
proving that the CVPM-MPC method is \gls{ISS}. We start with the definition of an \gls{ISS} Lyapunov function.
\begin{definition}[\gls{ISS} Lyapunov Function \cite{GoulartKerriganMaciejowski2006}]\label{def:ISS}
	Consider a continuous function $V: \gls{Xrci} \rightarrow \mathbb{R}$ and the functions $\alpha_1, \alpha_2, \alpha_3 \in \mathcal{K}_{\infty}$, $\gamma \in \mathcal{K}$. Then, $V$ is an ISS Lyapunov function for a system ${\bm{x}_{t+1}=\bm{f}(\bm{x}_t, \bm{w}_t)}$, where $\bm{f}$ is continuous,  with a robust positively invariant set \gls{Xrci} if $\alpha_1, \alpha_2, \alpha_3, \gamma$ exist, for all $k \geq 0$ and $\bm{x}_t \in \gls{Xrci}$, such that
	\vspace{-1mm}
	\begin{IEEEeqnarray}{l}
		\alpha_1(||\bm{x}_t||) \leq V(\bm{x}_t) \leq \alpha_2 (||\bm{x}_t||) \IEEEyessubnumber \label{eq:Vbound} \\
		V(\bm{f}(\bm{x}_{t},\bm{w}_t)) - V(\bm{x}_t) \leq - \alpha_3(||\bm{x}_t||) + \gamma(||\bm{w}_t||). \IEEEyessubnumber \IEEEeqnarraynumspace \label{eq:Vdec1}
	\end{IEEEeqnarray}
\end{definition}

\begin{lemma}[Sufficient \gls{ISS} Condition \cite{GoulartKerriganMaciejowski2006}]
	The origin of system $\bm{x}_{t+1}=\bm{f}(\bm{x}_t, \bm{w}_t)$ where $f$ is continuous is \gls{ISS} if an ISS Lyapunov function according to Definition~\ref{def:ISS} exists.
\end{lemma}

The following assumption is required.

\begin{assumption}
	\label{ass:cost}
	The weighting matrices of the stage cost are positive definite and symmetric, i.e.,  $\bm{Q} = \bm{Q}^\top \succ 0$ and $\bm{R} = \bm{R}^\top \succ 0$.
	The terminal cost weighting matrix \gls{terminalCost} is a solution of the 
	discrete-time algebraic Riccati equation.
\end{assumption}

Based on the robust control invariance of  \gls{case1Set} it can be shown that the origin is ISS for \gls{c1}.  

\begin{lemma}\label{lem:stab1}
	Let Assumptions \ref{ass:xP_origin}, \ref{ass:stabelA}, \ref{ass:rci} and  \ref{ass:cost} hold. Then, for ${\glspl{state}_t \in \gls{case1Set}}$, the origin of the closed-loop system~\eqref{eq:sys_cl} is ISS.
\end{lemma}

\begin{proof}
	The proof is given in Appendix~\ref{sec:appendixa}.
\end{proof}

Consecutively applying the safe case yields same behavior as robust MPC. 
If \gls{c1} is not applicable, i.e., $\glspl{state}_t \notin \gls{case1Set}$, it needs to be ensured that the system is still ISS.
For the inverse covariance matrix of the last predicted state $\glspl{state}_{\gls{Ncvpm}}$, the solution $\bs{S}$ of the discrete-time algebraic Riccati equation is used, i.e.,
\begin{align}
	(\gls{sysMat}-\gls{inputMat}\bs{K})^\top \bs{S} (\gls{sysMat}-\gls{inputMat}\bs{K}) + \gls{Sigmax}^{-1} = \bs{S}
\end{align}
where $\bs{K}$ is a control gain such that ${\glspl{state}_t \in \gls{case1Set} \;\Rightarrow\;  \glspl{state}_{t+1} \in \gls{case1Set}}$. 
It follows that
\begin{IEEEeqnarray}{c}\label{eq:costCase2}
	\glspl{SigmaX}^{-1} =\mathrm{diag}\lr{\gls{Sigmax}^{-1}, \cdots, \gls{Sigmax}^{-1}, \bs{S}}
\end{IEEEeqnarray}
describes the adapted inverse of the covariance matrix.

\begin{lemma}\label{lem:stab2}
	Let Assumptions \ref{ass:stabelA} and \ref{ass:cost} hold. For $\glspl{state}_t \notin \gls{case1Set}$, 
	the origin of the error dynamics of the error between the closed-loop system~\eqref{eq:sys_cl} and the optimization variable $\gls{int_xi}\in\gls{case1Set}$ is \gls{ISS} using the method from Section~\ref{sec:approx}. 
\end{lemma}

\begin{proof}
	The proof is given in Appendix~\ref{sec:appendixb}.
\end{proof}

Based on Lemma~\ref{lem:stab1} and Lemma~\ref{lem:stab2}, we can now formulate the stability theorem for CVPM-MPC.

\begin{theorem}\label{th:stabil}
	Let Assumptions \ref{ass:xP_origin}, \ref{ass:stabelA}, \ref{ass:rci} and \ref{ass:cost}  hold. The origin of the closed-loop system~\eqref{eq:sys_cl}, controlled by CVPM-MPC, is \gls{ISS}.
\end{theorem}

\begin{proof}
	 From Lemma~\ref{lem:stab2} we conclude that the state converges to the set \gls{case1Set}. Based on Lemma~\ref{lem:stab1} the origin of system~\eqref{eq:sys_cl} is \gls{ISS} for $\gls{state}\in \gls{case1Set}$. Finally, the origin is \gls{ISS} for all $\gls{state} \in \mathbb{R}^{\gls{dimx}}$.
\end{proof}

\subsection{Extension to Time-Variant Constraints}\label{sec:timevariant}

We have assumed that the state constraint set \gls{probStateSet} and the disturbance set \gls{distSet} are time-invariant. However, the method can also handle time-variant \gls{probStateSet} and \gls{distSet} during runtime without preprocessing. We discuss two different situations here. First, we discuss stability when the sets \gls{probStateSet} and \gls{distSet} change for a short time, and second, when the sets change permanently.  

\subsubsection{Short-term change} 
A change of  the sets  \gls{probStateSet} or \gls{distSet}  for one time step, i.e., an unexpected large disturbance, may lead to a violation of Assumptions~\ref{ass:xP_origin} and \ref{ass:rci}, and thus invalidate the stability guarantee in \gls{c1}.
A violation of Assumption~\ref{ass:distSubX} leads to $\gls{case1Set}=\emptyset$ and, therefore, \gls{c2} is applied.
Although recursive feasibility is preserved, stability of the origin can no longer be ensured.
As the temporary change subsides, all assumptions are again satisfied, resulting in stability.

\subsubsection{Permanent change}

It can be shown that, under Assumptions~\ref{ass:xP_origin}, \ref{ass:distSubX}, and \ref{ass:rci}, also a permanent change of the sets \gls{probStateSet} and \gls{distSet} does not affect the stability of the system. 

To meet Assumption \ref{ass:rci}, a precomputed terminal set \gls{terminalSet} can be utilized, which is computed for an initially large disturbance set \gls{distSet}. 
If the disturbance set changes during runtime, the assumptions still hold. 
A suitable choice for \gls{terminalSet} is a minimal robust invariant set. This, due to $\gls{terminalSet} \subseteq\gls{probStateSet}$, enables to choose the state constraint set \gls{probStateSet} rather small.

\section{Numerical Example}
\label{sec:results}
In the following, we discuss recursive feasibility and stability of the proposed method in simulation studies and demonstrate the capability of CVPM-MPC to handle time-variant uncertainty bounds.

\subsection{Simulation Setup}

We consider a discrete-time linear system. The system matrix, input matrix, and disturbance matrix are given by
\begin{IEEEeqnarray}{c}
	\gls{sysMat}=\begin{bmatrix}
		0.99&\text{-}0.02\\0.21&0.92
	\end{bmatrix},
	\gls{inputMat}=
	\begin{bmatrix}
		0.30\\0.06
	\end{bmatrix}, 
	\gls{distMat}=\begin{bmatrix}
		0.02&0.00\\0.01&0.19
	\end{bmatrix}. \IEEEeqnarraynumspace
	\label{eq:simSys}
\end{IEEEeqnarray}
The model describes a DC-DC converter, see \cite{Tan2015}, where the state $x_1$ stands for the current in a coil and the state $x_2$ is the voltage of a capacitor. 
The goal is to stabilize a voltage of \SI{3.3}{\volt}, yielding the reference 
$\glspl{state}_\text{ref} =  \begin{bmatrix}1.06&3.30\end{bmatrix}^\top$, and
$\glspl{input}_\text{ref} = 0.28.$
The method is adapted such that the reference state is stabilized.
The input is the duty cycle of a transistor, thus
$\gls{inputSet} = \left\lbrace \glspl{input} \;\middle|\; 0\leq \glspl{input} \leq 1	\right\rbrace. $
In the simulation, modeled and unmodeled disturbances are considered. 
The support of the modeled disturbance is
\begin{IEEEeqnarray}{c}
\gls{distSet} = \left\lbrace[w_1,w_2]^\top \middle|\; -0.2\leq w_1 \leq 0.2,-0.2\leq w_2 \leq 0.2 \right\rbrace \IEEEeqnarraynumspace
\end{IEEEeqnarray}
with covariance matrix $\gls{Sigmaw} = \operatorname{diag}(0.2,0.2)$.
The time-invariant state constraint set is chosen as
\begin{IEEEeqnarray}{c}
	\gls{probStateSet} = \left\lbrace[x_1,x_2]^\top \middle|\; 0\leq x_1 \leq 2,2.8\leq x_2 \leq 3.8 \right\rbrace. 
\end{IEEEeqnarray}
Note that state constraints, uncertainty, and system matrix satisfy Assumptions~\ref{ass:uncertainy}, \ref{ass:xP_origin} and \ref{ass:stabelA}.

The MPC employs a horizon of $\gls{Ncvpm}=10$ with sampling time $\Delta t = 0.1$ and  the weighting matrices are chosen as
${\gls{stateCost}=\operatorname{diag}(1,5)}$, and $\gls{inputCost}=1$. 
 The terminal cost \gls{terminalCost} is determined according to Assumption~\ref{ass:cost}. 
The computation of the polyhedra \gls{optInputSet} and \gls{case1Set} is done with the MPT3 toolbox  \cite{HercegEtalMorari2013}.

\subsection{Comparison of Probability Minimization Methods}\label{sec:comparison}

The minimum of the constraint violation probability  \eqref{eq:proboptim} is challenging to compute. Therefore, two methods to approximate the probability are introduced. First, in Section~\ref{sec:numProb}, a numeric computation of the probability using a Monte Carlo method, and in Section~\ref{sec:approx}, an approximation of the probability utilizing a quadratic program.

\setlength\fwidth{0.34\columnwidth}
\setlength\fheight{0.25\columnwidth}
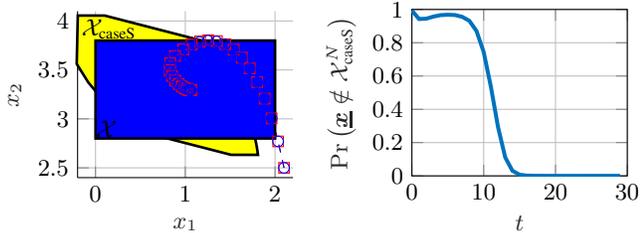
\begin{figure}[ht]
	\begin{subfigure}[b]{0.49\columnwidth}
		\centering
%
\definecolor{mycolor1}{rgb}{1.00000,1.00000,0.00000}%
\begin{tikzpicture}

\begin{axis}[%
width=0.951\fwidth,
height=\fheight,
at={(0\fwidth,0\fheight)},
scale only axis,
clip=false,
xmin=-0.2,
xmax=2.2,
xlabel style={font=\color{white!15!black}},
xlabel={$x_1$},
ymin=2.4,
ymax=4.1,
ylabel style={font=\color{white!15!black}},
ylabel={$x_2$},
axis background/.style={fill=white},
axis x line*=bottom,
axis y line*=left,
xmajorgrids,
ymajorgrids
]

\addplot[area legend, line width=1.0pt, draw=black, fill=mycolor1, fill opacity=0.1, forget plot]
table[row sep=crcr] {%
x	y\\
0.134153566750988	3.1360846676508\\
0.35442335386681	2.96416884321687\\
0.578218382127193	2.85342644368486\\
1.5099824872523	2.63300025807189\\
1.80984044152877	2.63256989800012\\
1.80464708886031	2.66788793465856\\
1.79913756981705	2.70111657698637\\
1.79525010623707	2.72152791515532\\
1.78481041656877	2.77142948432896\\
1.77950827117147	2.79163442619845\\
1.59594972365424	3.48575561739073\\
1.53986739076193	3.56799189549155\\
1.45274194936694	3.66391292076137\\
1.36355051977083	3.73352491045922\\
1.272931652933	3.77836661598995\\
0.102251303297187	4.0553128973748\\
-0.197606650979284	4.05574325744657\\
-0.209529897468885	3.56142324030753\\
-0.0810139949234021	3.37297406287401\\
}--cycle;
\addplot [color=red, dashed, mark=square, mark options={solid, red}, forget plot]
  table[row sep=crcr]{%
2.1	2.5\\
2.03405040858589	2.77005906860306\\
1.96178752337101	3.00526447709526\\
1.88406718685466	3.20687269796137\\
1.80171264871458	3.37623228708515\\
1.71811361144741	3.51533668729007\\
1.63011380256256	3.62535561534451\\
1.53878507323635	3.70760580779669\\
1.44554874842699	3.76362247335322\\
1.35197591143166	3.79517701322855\\
1.25753334891192	3.80380197977956\\
1.16510621934725	3.79155008606939\\
1.07067897146945	3.75945820161305\\
0.977285140477171	3.70914595229306\\
0.88536013202873	3.64221785409323\\
0.830998751408305	3.56803249994335\\
0.827709615484336	3.49828169511005\\
0.852054292104382	3.43874380737547\\
0.888412392446739	3.39135474038609\\
0.927094020873525	3.35578142706971\\
0.962665077350401	3.33051702150517\\
0.992574557005917	3.31359757085209\\
1.01609748342667	3.30303601351337\\
1.03356520344659	3.29705645706325\\
1.045835681187	3.29419548831744\\
1.05395154485334	3.29332111766262\\
1.05893666974788	3.29360560058491\\
1.06168978450142	3.29447671423836\\
1.06294171491786	3.29556300477267\\
1.06325145146039	3.29664197608528\\
};
\addplot [color=blue, dashed, mark=o, mark options={solid, blue}, forget plot]
  table[row sep=crcr]{%
2.1	2.5\\
2.03405040858589	2.77005906860306\\
1.96178752337101	3.00526447709526\\
1.88406718685466	3.20687269796137\\
1.80171264871458	3.37623228708515\\
1.7155124411161	3.5147697897168\\
1.62621859891856	3.62397624817769\\
1.53454520935015	3.70539434119405\\
1.44116727504564	3.76060617911599\\
1.34671987380285	3.79122177392121\\
1.25179759800504	3.79886819783304\\
1.15695425637208	3.78517943960782\\
1.06270282059131	3.75178696308252\\
0.969515598900512	3.70031096832031\\
0.877824624491126	3.63235235299972\\
0.830385622013639	3.55871801899044\\
0.830860817792323	3.4903090649144\\
0.856860554295505	3.43238426385699\\
0.893561561639506	3.38657163592027\\
0.931847061349282	3.35237787901674\\
0.966672027233963	3.32823275462928\\
0.995732517016255	3.31216774499931\\
1.01844701807681	3.30222314109491\\
1.0352188994899	3.29666436485871\\
1.04693190925115	3.29407212365343\\
1.05462687566221	3.29335508370698\\
1.05931135162338	3.29371973378924\\
1.06186182673439	3.29462080519566\\
1.06298623078514	3.29570688057394\\
1.06322288287679	3.29676955375848\\
};

\addplot[area legend, line width=1.0pt, draw=black, fill=blue, fill opacity=0.1, forget plot]
table[row sep=crcr] {%
x	y\\
0	2.8\\
2	2.8\\
2	3.8\\
0	3.8\\
}--cycle;
\node[right, align=left]
at (axis cs:-0.09,2.9) {$\mathcal{X}$};
\node[right, align=left]
at (axis cs:-0.25,3.92) {$\mathcal{X}_\textrm{caseS}$};
\end{axis}
\end{tikzpicture}%
	\end{subfigure}
	\begin{subfigure}[b]{0.49\columnwidth}
		\centering
%
\definecolor{mycolor1}{rgb}{0.00000,0.44700,0.74100}%
\begin{tikzpicture}

\begin{axis}[%
width=0.951\fwidth,
height=\fheight,
at={(0\fwidth,0\fheight)},
scale only axis,
xmin=0,
xmax=30,
xlabel style={font=\color{white!15!black}},
xlabel={$t$},
ymin=0,
ymax=1,
ylabel style={font=\color{white!15!black}},
ylabel={$\Pr\left(\glspl{states}\notin \gls{case1Set}^N\right)$},
axis background/.style={fill=white},
xmajorgrids,
ymajorgrids
]
\addplot [color=mycolor1, line width=1.5pt, forget plot]
  table[row sep=crcr]{%
0	0.993549999999999\\
1	0.942499999999999\\
2	0.944490000000002\\
3	0.957080000000001\\
4	0.96583\\
5	0.968160000000001\\
6	0.966470000000001\\
7	0.955839999999998\\
8	0.931760000000001\\
9	0.87032\\
10	0.743659999999998\\
11	0.53584\\
12	0.292370000000002\\
13	0.110959999999999\\
14	0.030619999999999\\
15	0.00707999999999842\\
16	0.0018199999999986\\
17	0.000859999999999417\\
19	0.000260000000000815\\
29	0.000350000000000961\\
};
\end{axis}
\end{tikzpicture}%
	\end{subfigure}
	\caption{
		Comparison of  probability minimization methods;
		Left: trajectory of Monte Carlo Integration (red), Quadratic Program approximation  (blue, Sec. \ref{sec:approx}); Right: constraint violation probability
	}
	\label{fig:compare}
\end{figure}

As shown on the left in Figure~\ref{fig:compare}, the trajectories that result from applying both methods are almost identical. The initial value is not in \gls{probStateSet} leading to a high probability of constraint violation. Applying the proposed method leads to a convergence to the set \gls{probStateSet} and a decrease in the constraint violation probability.  
On the right side in Figure~\ref{fig:compare}, the constraint violation probability for the approach from Section~\ref{sec:approx} is given. 
Initially the constraint violation probability is almost 1 and the minimization allows to approach the set \gls{case1Set}.  
We observe that \gls{c2} can deal with otherwise infeasible initial states and find a trajectory that converges to \gls{case1Set}. 

The average computation time for one step of the approximation introduced in Section~\ref{sec:approx} is  \SI{60}{\milli\second} on a standard computer while the average computation time of one step of the sampling method is  \SI{4}{\minute} on a computer with an Intel Xeon E5-2630. 
The sampling-based approach uses $N_\text{s}=10^5$ randomly generated samples in each iteration distributed according to \eqref{eq:statesequenceGauss}.

\subsection{Performance with Modeled and Unmodeled Disturbances }

The following simulation shows convergence to a reference when starting with a non-zero constraint violation probability. 
At time step $t=50$, an unmodeled disturbance affects the system for one time step, which is handled by the CVPM-MPC method. An unmodeled disturbance may be interpreted as an increase of \gls{distSet} for one time step or as $\gls{dist} \notin \gls{distSet}$.
Figure~\ref{fig:sim} illustrates the simulation results.

\setlength\fwidth{0.38\columnwidth}
\setlength\fheight{0.2\columnwidth}
\begin{figure}[ht]
	\begin{subfigure}[b]{0.49\columnwidth}
		\centering
%
\definecolor{mycolor1}{rgb}{1.00000,1.00000,0.00000}%
\begin{tikzpicture}

\begin{axis}[%
width=0.951\fwidth,
height=\fheight,
at={(0\fwidth,0\fheight)},
scale only axis,
xmin=-0.32,
xmax=2.1105,
xlabel style={font=\color{white!15!black}},
xlabel={$x_1$},
ymin=0,
ymax=4.0557,
ylabel style={font=\color{white!15!black}},
ylabel={$x_2$},
axis background/.style={fill=white},
axis x line*=bottom,
axis y line*=left,
xmajorgrids,
ymajorgrids
]

\addplot[area legend, line width=1.0pt, draw=black, fill=mycolor1, fill opacity=0.1, forget plot]
table[row sep=crcr] {%
x	y\\
0.134153566750988	3.1360846676508\\
0.35442335386681	2.96416884321687\\
0.578218382127193	2.85342644368486\\
1.5099824872523	2.63300025807189\\
1.80984044152877	2.63256989800012\\
1.80464708886031	2.66788793465856\\
1.79913756981705	2.70111657698637\\
1.79525010623707	2.72152791515532\\
1.78481041656877	2.77142948432896\\
1.77950827117147	2.79163442619845\\
1.59594972365424	3.48575561739073\\
1.53986739076193	3.56799189549155\\
1.45274194936694	3.66391292076137\\
1.36355051977083	3.73352491045922\\
1.272931652933	3.77836661598995\\
0.102251303297187	4.0553128973748\\
-0.197606650979284	4.05574325744657\\
-0.209529897468885	3.56142324030753\\
-0.0810139949234021	3.37297406287401\\
}--cycle;

\addplot[area legend, line width=1.0pt, draw=black, fill=blue, fill opacity=0.1, forget plot]
table[row sep=crcr] {%
x	y\\
0	2.8\\
2	2.8\\
2	3.8\\
0	3.8\\
}--cycle;
\addplot [color=black, dashed, forget plot]
  table[row sep=crcr]{%
0	0\\
0.302159508034426	0.0617981268391321\\
0.598288131110265	0.202513381483078\\
0.89154761822659	0.40971205446378\\
1.1770348912033	0.643227415620588\\
1.45906636150742	0.906725206975057\\
1.73396999655578	1.21013537604429\\
1.84022218581284	1.5371957239668\\
1.81887576897901	1.82359375028822\\
1.80896477334991	2.36965247368238\\
1.80781075714095	2.58508693391678\\
1.74323558005746	2.79400840147881\\
1.67225041555706	2.97118005094573\\
1.59349058653729	3.09874025486848\\
1.51862319458999	3.18738145091433\\
1.43509710690916	3.29654049643099\\
1.34833113165786	3.35327053300831\\
1.26215485927829	3.41148076416014\\
1.17880291412328	3.45635168362968\\
1.09099618864042	3.44355494743281\\
1.02588594500601	3.41688179183648\\
0.994894967629419	3.41304070092693\\
0.97637430395296	3.39733500408661\\
0.971911397753536	3.36910989146544\\
0.986288823302497	3.34997684034178\\
1.00152866703761	3.30689325585242\\
1.02059761178408	3.27741253673471\\
1.04581760154586	3.24946552002083\\
1.06768587679838	3.27832376673666\\
1.07202220606639	3.29449910068388\\
1.0682261387292	3.2909141914185\\
1.06687819042735	3.3193317231756\\
1.05902880891664	3.33229333971067\\
1.04460609285406	3.33462207853674\\
1.03591463708998	3.32133070136937\\
1.03824039644137	3.310421886646\\
1.04310197408822	3.32493300331907\\
1.04017671552301	3.34371545663413\\
1.03541229947407	3.31257493952852\\
1.03894892963587	3.3194363724176\\
1.05311804485807	3.29340649891576\\
1.05720141727381	3.30415319457862\\
1.05442316876443	3.31474832146378\\
1.05121191952652	3.297528612743\\
1.0546893340834	3.31168917833214\\
};
\addplot[only marks, mark=*, mark options={}, mark size=1.2500pt, color=black!20!green, fill=black!20!green, forget plot] table[row sep=crcr]{%
x	y\\
1.74323558005746	2.79400840147881\\
1.67225041555706	2.97118005094573\\
1.59349058653729	3.09874025486848\\
1.51862319458999	3.18738145091433\\
1.43509710690916	3.29654049643099\\
1.34833113165786	3.35327053300831\\
1.26215485927829	3.41148076416014\\
1.17880291412328	3.45635168362968\\
1.09099618864042	3.44355494743281\\
1.02588594500601	3.41688179183648\\
0.994894967629419	3.41304070092693\\
0.97637430395296	3.39733500408661\\
0.971911397753536	3.36910989146544\\
0.986288823302497	3.34997684034178\\
1.00152866703761	3.30689325585242\\
1.02059761178408	3.27741253673471\\
1.04581760154586	3.24946552002083\\
1.06768587679838	3.27832376673666\\
1.07202220606639	3.29449910068388\\
1.0682261387292	3.2909141914185\\
1.06687819042735	3.3193317231756\\
1.05902880891664	3.33229333971067\\
1.04460609285405	3.33462207853674\\
1.03591463708998	3.32133070136937\\
1.03824039644137	3.310421886646\\
1.04310197408822	3.32493300331907\\
1.04017671552301	3.34371545663413\\
1.03541229947407	3.31257493952852\\
1.03894892963587	3.3194363724176\\
1.04154480511282	3.31483858735891\\
1.04315149216128	3.31146609554862\\
1.04681106142935	3.30549436673506\\
1.05311804485807	3.29340649891576\\
1.05720141727381	3.30415319457862\\
1.05442316876443	3.31474832146378\\
1.05121191952652	3.297528612743\\
1.0546893340834	3.31168917833214\\
};
\addplot[only marks, mark=*, mark options={}, mark size=1.2500pt, color=black!20!red, fill=black!20!red, forget plot] table[row sep=crcr]{%
x	y\\
0	0\\
0.302159508034425	0.0617981268391322\\
0.598288131110265	0.202513381483078\\
0.89154761822659	0.409712054463781\\
1.1770348912033	0.643227415620588\\
1.45906636150742	0.906725206975057\\
1.73396999655578	1.21013537604429\\
1.84022218581284	1.5371957239668\\
1.81887576897901	1.82359375028822\\
1.81383826516458	2.11227886980206\\
1.80896477334991	2.36965247368238\\
1.80781075714095	2.58508693391678\\
};
\addplot[only marks, mark=*, mark options={}, mark size=1.5000pt, color=green, fill=green, forget plot] table[row sep=crcr]{%
x	y\\
1.0546893340834	3.31168917833214\\
};
\end{axis}
\end{tikzpicture}%
	\end{subfigure}
	\begin{subfigure}[b]{0.49\columnwidth}
		\centering
%
\definecolor{mycolor1}{rgb}{1.00000,1.00000,0.00000}%
\begin{tikzpicture}

\begin{axis}[%
width=0.951\fwidth,
height=\fheight,
at={(0\fwidth,0\fheight)},
scale only axis,
xmin=-0.32,
xmax=2.1105,
xlabel style={font=\color{white!15!black}},
xlabel={$x_1$},
ymin=0,
ymax=4.0557,
ylabel style={font=\color{white!15!black}},
ylabel={$x_2$},
axis background/.style={fill=white},
axis x line*=bottom,
axis y line*=left,
xmajorgrids,
ymajorgrids
]

\addplot[area legend, line width=1.0pt, draw=black, fill=mycolor1, fill opacity=0.1, forget plot]
table[row sep=crcr] {%
x	y\\
0.134153566750988	3.1360846676508\\
0.35442335386681	2.96416884321687\\
0.578218382127193	2.85342644368486\\
1.5099824872523	2.63300025807189\\
1.80984044152877	2.63256989800012\\
1.80464708886031	2.66788793465856\\
1.79913756981705	2.70111657698637\\
1.79525010623707	2.72152791515532\\
1.78481041656877	2.77142948432896\\
1.77950827117147	2.79163442619845\\
1.59594972365424	3.48575561739073\\
1.53986739076193	3.56799189549155\\
1.45274194936694	3.66391292076137\\
1.36355051977083	3.73352491045922\\
1.272931652933	3.77836661598995\\
0.102251303297187	4.0553128973748\\
-0.197606650979284	4.05574325744657\\
-0.209529897468885	3.56142324030753\\
-0.0810139949234021	3.37297406287401\\
}--cycle;

\addplot[area legend, line width=1.0pt, draw=black, fill=blue, fill opacity=0.1, forget plot]
table[row sep=crcr] {%
x	y\\
0	2.8\\
2	2.8\\
2	3.8\\
0	3.8\\
}--cycle;
\addplot [color=black, dashed, forget plot]
  table[row sep=crcr]{%
0	0\\
0.302159508034426	0.0617981268391321\\
0.598288131110265	0.202513381483078\\
0.89154761822659	0.40971205446378\\
1.1770348912033	0.643227415620588\\
1.45906636150742	0.906725206975057\\
1.73396999655578	1.21013537604429\\
1.84022218581284	1.5371957239668\\
1.81887576897901	1.82359375028822\\
1.80896477334991	2.36965247368238\\
1.80781075714095	2.58508693391678\\
1.74323558005746	2.79400840147881\\
1.67225041555706	2.97118005094573\\
1.59349058653729	3.09874025486848\\
1.51862319458999	3.18738145091433\\
1.43509710690916	3.29654049643099\\
1.34833113165786	3.35327053300831\\
1.26215485927829	3.41148076416014\\
1.17880291412328	3.45635168362968\\
1.09099618864042	3.44355494743281\\
1.02588594500601	3.41688179183648\\
0.994894967629419	3.41304070092693\\
0.97637430395296	3.39733500408661\\
0.971911397753536	3.36910989146544\\
0.986288823302497	3.34997684034178\\
1.00152866703761	3.30689325585242\\
1.02059761178408	3.27741253673471\\
1.04581760154586	3.24946552002083\\
1.06768587679838	3.27832376673666\\
1.07202220606639	3.29449910068388\\
1.0682261387292	3.2909141914185\\
1.06687819042735	3.3193317231756\\
1.05902880891664	3.33229333971067\\
1.04460609285406	3.33462207853674\\
1.03591463708998	3.32133070136937\\
1.03824039644137	3.310421886646\\
1.04310197408822	3.32493300331907\\
1.04017671552301	3.34371545663413\\
1.03541229947407	3.31257493952852\\
1.03894892963587	3.3194363724176\\
1.05311804485807	3.29340649891576\\
1.05720141727381	3.30415319457862\\
1.05442316876443	3.31474832146378\\
1.05121191952652	3.297528612743\\
1.0546893340834	3.31168917833214\\
1.05521027120112	0\\
1.34855892496155	0.285169002928305\\
1.63811585359401	0.615391512080075\\
1.88191491353661	0.981585851423363\\
1.85046667936364	1.32446354324227\\
1.83167360770015	1.62176026259068\\
1.81545655922623	1.91089024478724\\
1.80884313664282	2.18552717170239\\
1.80893324226243	2.44115871883341\\
1.74770747670143	2.62565435114308\\
1.68251952061634	2.79163518874005\\
1.60891192523913	2.93040994531415\\
1.53717089887484	3.05037259738182\\
1.46329787113222	3.13199426945304\\
1.38603956361167	3.19603467100874\\
1.3013510738131	3.27871936773351\\
1.21769085617751	3.29987638244087\\
1.14952249438572	3.33145690392216\\
1.09847137541897	3.34026784771246\\
1.06443718038617	3.35728563271573\\
1.04389567110858	3.34991006331849\\
1.03370869414388	3.34349778385723\\
1.02470345735264	3.34226406783169\\
1.02202897171154	3.32431272657301\\
1.02936499671478	3.33198064568645\\
1.03435989438551	3.30900226402709\\
1.04169241913111	3.2791245994165\\
1.05696978654304	3.28823525327221\\
1.06510973776219	3.27151044444399\\
1.07602137653147	3.27117915675244\\
1.07790980712052	3.27449137652551\\
1.07871328877286	3.29913435627978\\
1.07007568261212	3.3008842684265\\
1.06343158890739	3.30370962618513\\
1.06071327515988	3.27767832646664\\
1.06931743225908	3.27207356165506\\
1.08366055061698	3.25371221645347\\
1.08763653733561	3.27322150598769\\
1.08654597943095	3.29784409818519\\
1.07772158462324	3.28792063599548\\
1.07351630358667	3.3171191593183\\
1.06352331470549	3.29391129061269\\
1.06157836542434	3.27475900645736\\
1.06813379957349	3.26915967897562\\
1.07671625339085	3.26837987902781\\
1.07770892241046	3.30250765557585\\
1.06864250184062	3.31234520316994\\
1.0605801827738	3.28763610680743\\
1.0655279824509	3.27336194698484\\
};
\addplot[only marks, mark=*, mark options={}, mark size=1.2500pt, color=black!20!green, fill=black!20!green, forget plot] table[row sep=crcr]{%
x	y\\
1.74323558005746	2.79400840147881\\
1.67225041555706	2.97118005094573\\
1.59349058653729	3.09874025486848\\
1.51862319458999	3.18738145091433\\
1.43509710690916	3.29654049643099\\
1.34833113165786	3.35327053300831\\
1.26215485927829	3.41148076416014\\
1.17880291412328	3.45635168362968\\
1.09099618864042	3.44355494743281\\
1.02588594500601	3.41688179183648\\
0.994894967629419	3.41304070092693\\
0.97637430395296	3.39733500408661\\
0.971911397753536	3.36910989146544\\
0.986288823302497	3.34997684034178\\
1.00152866703761	3.30689325585242\\
1.02059761178408	3.27741253673471\\
1.04581760154586	3.24946552002083\\
1.06768587679838	3.27832376673666\\
1.07202220606639	3.29449910068388\\
1.0682261387292	3.2909141914185\\
1.06687819042735	3.3193317231756\\
1.05902880891664	3.33229333971067\\
1.04460609285405	3.33462207853674\\
1.03591463708998	3.32133070136937\\
1.03824039644137	3.310421886646\\
1.04310197408822	3.32493300331907\\
1.04017671552301	3.34371545663413\\
1.03541229947407	3.31257493952852\\
1.03894892963587	3.3194363724176\\
1.04154480511282	3.31483858735891\\
1.04315149216128	3.31146609554862\\
1.04681106142935	3.30549436673506\\
1.05311804485807	3.29340649891576\\
1.05720141727381	3.30415319457862\\
1.05442316876443	3.31474832146378\\
1.05121191952652	3.297528612743\\
1.0546893340834	3.31168917833214\\
1.68251952061634	2.79163518874005\\
1.60891192523913	2.93040994531415\\
1.53717089887484	3.05037259738182\\
1.46329787113222	3.13199426945304\\
1.38603956361166	3.19603467100874\\
1.3013510738131	3.27871936773351\\
1.21769085617751	3.29987638244087\\
1.14952249438572	3.33145690392216\\
1.09847137541897	3.34026784771246\\
1.06443718038617	3.35728563271573\\
1.04389567110858	3.34991006331849\\
1.03370869414387	3.34349778385723\\
1.02624380487775	3.34275705996852\\
1.02470345735264	3.34226406783169\\
1.02202897171154	3.32431272657301\\
1.02936499671478	3.33198064568645\\
1.03435989438551	3.30900226402709\\
1.04169241913111	3.2791245994165\\
1.05696978654304	3.28823525327221\\
1.06510973776219	3.27151044444399\\
1.07602137653147	3.27117915675244\\
1.07790980712052	3.27449137652551\\
1.07871328877286	3.29913435627978\\
1.07007568261212	3.3008842684265\\
1.06343158890739	3.30370962618513\\
1.06071327515988	3.27767832646664\\
1.06931743225908	3.27207356165506\\
1.07725524712085	3.26212967485484\\
1.08366055061698	3.25371221645347\\
1.08763653733561	3.27322150598769\\
1.08654597943095	3.29784409818519\\
1.07772158462324	3.28792063599548\\
1.07351630358667	3.3171191593183\\
1.06352331470549	3.29391129061269\\
1.06157836542434	3.27475900645736\\
1.06813379957349	3.26915967897562\\
1.07671625339085	3.26837987902781\\
1.07770892241046	3.30250765557585\\
1.06864250184062	3.31234520316994\\
1.0605801827738	3.28763610680743\\
1.0655279824509	3.27336194698484\\
};
\addplot[only marks, mark=*, mark options={}, mark size=1.2500pt, color=black!20!red, fill=black!20!red, forget plot] table[row sep=crcr]{%
x	y\\
0	0\\
0.302159508034425	0.0617981268391322\\
0.598288131110265	0.202513381483078\\
0.89154761822659	0.409712054463781\\
1.1770348912033	0.643227415620588\\
1.45906636150742	0.906725206975057\\
1.73396999655578	1.21013537604429\\
1.84022218581284	1.5371957239668\\
1.81887576897901	1.82359375028822\\
1.81383826516458	2.11227886980206\\
1.80896477334991	2.36965247368238\\
1.80781075714095	2.58508693391678\\
1.05521027120112	0\\
1.34855892496155	0.285169002928306\\
1.63811585359401	0.615391512080075\\
1.88191491353661	0.981585851423363\\
1.85046667936364	1.32446354324227\\
1.83167360770015	1.62176026259068\\
1.81545655922623	1.91089024478724\\
1.80884313664282	2.18552717170239\\
1.80893324226243	2.44115871883341\\
1.74770747670143	2.62565435114308\\
};
\addplot[only marks, mark=*, mark options={}, mark size=1.5000pt, color=green, fill=green, forget plot] table[row sep=crcr]{%
x	y\\
1.0655279824509	3.27336194698484\\
};
\end{axis}
\end{tikzpicture}%
	\end{subfigure}
	\caption{Simulation of CVPM-MPC with an unmodeled disturbance: 
		Left: convergence to the set $\mathcal{X}$; Right: convergence to the set $\mathcal{X}$ after unmodeled disturbance.
		(green: \gls{c1b}; red: \gls{c2b}; bright marker: current state)} 
	\label{fig:sim}
\end{figure}
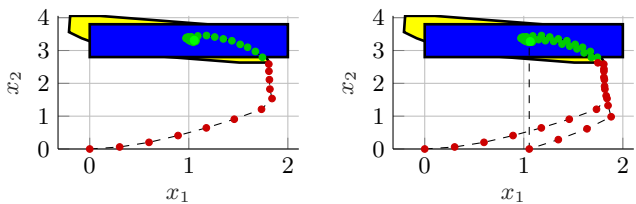

The set \gls{probStateSet} is indicated by the blue box. The set $\gls{case1Set}$ is marked in yellow. If the system state is in \gls{case1Set}, \gls{c1} is applicable. 
The initial state does not allow zero constraint violation probability in the next step, i.e., $\glspl{state}_0 \notin \gls{case1Set}$; therefore, \gls{c2} is required, indicated by the red dot in Figure~\ref{fig:sim}. 

Applying the CVPM-MPC procedure for \gls{c2} moves the system state into the set \gls{case1Set}, as seen on the left side in
Figure~\ref{fig:sim}. 
In \gls{case1Set} the control input is determined based on \gls{c1}, as indicated by the green dots since it is possible to reach \gls{probStateSet} in the next step.
The subsequent steps with \gls{c1} move the system state towards the origin. 
Not that in this simulation, mostly  \gls{c2} is active whereas  in application the standard situation is applying \gls{c1} and only switching to \gls{c2} when unexpected disturbances occur.

At time step $t=50$, an unmodeled disturbance occurs, which moves the system state outside of \gls{case1Set}, as illustrated on the right side in 
Figure~\ref{fig:sim}. Note that input-to-state stability is not guaranteed in this step as the uncertainty bound increased which violates Assumption~\ref{ass:uncertainy}. Similar to the initial simulation state, \gls{c2} is required because it is not possible to reach the constraint set \gls{probStateSet} in the next step. By switching from \gls{c1} to \gls{c2}, recursive feasibility is maintained. Afterwards, the CVPM-MPC method {steers the system state back to \gls{case1Set}.

\section{Discussion}
\label{sec:discussion}

In contrast to the \gls{cvpm} method presented in \cite{BruedigamEtalLeibold2021c} and \cite{Fink2022}, the \gls{cvpm} approach proposed in this work is more general. Here, we consider general linear constraints for a system with additive uncertainty, whereas \cite{BruedigamEtalLeibold2021c,Fink2022} 
are motivated by a vehicle collision avoidance scenario utilizing two dynamics and norm constraints.
The proposed approach significantly extends the possible applications to all linear or linearized systems where constraints are linear or where constraints may be linearized. 
The stability of the origin with general linear constraints is achieved with a robust control invariant set of initial states for \gls{c1} and Assumption~\ref{ass:rci} on the terminal constraint.he workspace of a robot.

In Section~\ref{sec:stability}, stability is discussed. As seen in the simulation example, the \gls{cvpm} method is capable of remaining feasible even in the presence of unmodeled disturbances, but at the cost of a constraint violation probability close to 1. The stability results may not hold in case of unmodeled disturbances, as a bounded uncertainty is assumed in the proofs (Assumption~\ref{ass:uncertainy}). 
The proposed method, however, allows updating the assumed uncertainty bound and state constraint. 
Therefore, the stability proof becomes valid again for an updated set \gls{distSet}  or \gls{probStateSet} as long as Assumptions~\ref{ass:xP_origin}, \ref{ass:distSubX}, and \ref{ass:rci} hold. 
If the uncertainty bound is not known initially, a conservative guess may be chosen and then the bound may be tightened over time, based on recorded data. The potential short loss of a stability guarantee is acceptable, however, as the main focus of this work is the minimization of constraint violation probability. Note that recursive feasibility remains guaranteed even for unmodeled disturbances.

If only SMPC is applied to the scenario shown in Fig.~\ref{fig:sim}, an 8~\% chance of violating the $x_1<2$ constraint is observed, while CVPM avoids constraint violation robustly.
In contrast to SMPC, \gls{c2} does not optimize a control objective but focuses only on the constraint by minimizing the probability of constraint violation. 
Only when the measured state, which is initial value for the prediction, is in \gls{case1Set}, the control objective is again taken into account.
We assume to have applications where \gls{c1} is active almost all the time while the \gls{c2} tackles rare problems with safety and is active if initial values are not safe or unmodeled disturbances disturbances occur.

In contrast to RMPC, we consider the disturbance set \gls{distSet} as a tuning parameter. 
Conservative assumptions with large \gls{distSet} results in conservative policies, whereas a small \gls{distSet} leads to a more optimistic but also more risky behavior.

\section{Conclusion}
\label{sec:conclusion}

The proposed CVPM-MPC method provides an MPC approach that combines the advantages of robust and stochastic MPC. The ability of CVPM-MPC to cope with time-variant constraints and uncertainty bounds provides a significant benefit for safety-critical systems. Recursive feasibility is guaranteed and stability is ensured under assumptions on the state constraint and disturbance support. 

CVPM-MPC is suitable for linear and linearized systems, enabling the use in applications such as quadcopter control or automated vehicles. 
Furthermore, the proposed CVPM-MPC method may be extended to consider probabilistic constraints and robust state constraints simultaneously. This extension would allow practitioners to employ robust constraints where possible and necessary as well as probabilistic CVPM constraints if suitable.

Whereas robust methods guarantee safety for predictable events, unpredictable environmental changes are not covered. This is especially complex if ethical concerns are relevant for applications, e.g., how autonomous systems should behave if collision avoidance cannot be guaranteed. 
CVPM-MPC provides a novel way to handle such scenarios and ethical issues.

\appendices

\section{Proof of Lemma~\ref{lem:rci}}
\label{sec:appendic}
In this section, we use the following notation.
The input sequence $\glspl{inputs}_{t}$ obtained at time step $t$ until prediction step ${t+N-1}$ yields the state sequence $\glspl{states}_{t}$, where these sequences are defined as 
\begin{IEEEeqnarray}{c}\label{eq:notation}
	\glspl{inputs}_{t} = (\bm{u}_{t|t}, ..., \bm{u}_{t+N-1|t}) \text{ and } 
	\glspl{states}_{t} = (\bm{x}_{t+1|t}, ..., \bm{x}_{t+N|t}).
	\IEEEeqnarraynumspace
\end{IEEEeqnarray}
The initial state for the state sequence is given as  $\bm{x}_{t|t} = \bm{x}_t$. 
Predictions made at time step $t+1$ are denoted
\begin{align}\label{eq:notation2}
	\glspl{state}_{t+1|t+1} &= \gls{sysMat}\glspl{state}_{t|t}+\gls{inputMat}\glspl{input}_{t|t} +\gls{distMat}\gls{dist} 
	= \glspl{state}_{t+1|t} +\gls{distMat}\gls{dist},
\end{align}
depending on the uncertainty \gls{dist} at time step $t$.
Optimal trajectories for the states and inputs are denoted by $\glspl{inputs}_{t}^*$ and  $\glspl{states}_{t}^*$. 

\begin{proof}
	We first show that a subsequent state sequence is in the zero violation set $ \gls{probStatesSet} \ominus \lr{\glspl{liftedDistMat}\circ \gls{distSet}^{\gls{Ncvpm}}} $  if the previous state sequence is also in the zero violation set. 
	
	Let the predicted state sequence $(\glspl{state}_{t+1|t} ... \glspl{state}_{t+\gls{Nmpc}|t})$, based on the initial state $\glspl{state}_t\in\gls{case1Set}$,  be in  $ \gls{probStatesSet} \ominus \lr{\glspl{liftedDistMat}\circ \gls{distSet}^{\gls{Ncvpm}}} $. 
	It follows that 
	\begin{IEEEeqnarray}{c}
	\glspl{state}_{t+i|t}  \in
	\gls{probStateSet} \ominus  \gls{sysMat}^{i-1}\gls{distMat}\circ\gls{distSet} \ominus  
	\gls{sysMat}^{i-2}\gls{distMat}\circ\gls{distSet} \ominus ... \ominus \gls{distMat}\circ\gls{distSet} . 
	\IEEEeqnarraynumspace
	\end{IEEEeqnarray} 
	Then, the subsequent state sequence intial state ${\glspl{state}_{t+1|t+1} = {\glspl{state}_{t+1|t}+\gls{distMat}\gls{dist}}}$ is  affected by the disturbance~\gls{dist}. 
	All states in the candidate state sequence are affected by the propagation of the disturbance yielding 
	\begin{IEEEeqnarray}{c}\label{eq:subseqSeq}
	\begin{bmatrix}
	\glspl{state}_{t+2|t+1} \\
	: \\
	\glspl{state}_{t+N|t+1} \\
	\glspl{state}_{t+N+1|t+1} 
	\end{bmatrix} = 
	\begin{bmatrix}
	\glspl{state}_{t+2|t} \\
	: \\
	\glspl{state}_{t+N|t} \\
	\gls{sysMat}\glspl{state}_{t+N|t}+\gls{inputMat}\glspl{input}
	\end{bmatrix} + 
	\begin{bmatrix}
	\gls{sysMat}\gls{distMat} \\
	: \\
	\gls{sysMat}^{\gls{Ncvpm}-1} \gls{distMat} \\
	\gls{sysMat}^{\gls{Ncvpm}} \gls{distMat} 
	\end{bmatrix} \glspl{dist}_t. \IEEEeqnarraynumspace
	\end{IEEEeqnarray}
	For a disturbance $\glspl{dist}_t\in\gls{distSet}$ the $t+i$-th state $\glspl{state}_{t+i|t+1}$ is in
	\begin{subequations}
	\begin{align}
	 &\gls{probStateSet} \ominus  \gls{sysMat}^{i-1}\gls{distMat}\circ\gls{distSet} \ominus  ... \ominus \gls{distMat}\circ\gls{distSet}  \oplus \gls{sysMat}^{i-1} \gls{distMat} \circ \gls{distSet} \\
	=& \gls{probStateSet} \ominus  \gls{sysMat}^{i-2}\gls{distMat}\circ\gls{distSet} \ominus  ... \ominus \gls{distMat}\circ\gls{distSet}  \qquad  \forall i \in \mathbb{I}_{2,N}.
	\end{align}
	\end{subequations}	
	Next, we show that the terminal state does not leave the terminal set. 	The state $\glspl{state}_{t+N|t}$ is in the terminal set \gls{terminalSet}. Based on Assumption~\ref{ass:rci}, it follows that an input 
	$\glspl{input}$
	exists such that the terminal state of the subsequent sequence is also in the terminal set, i.e., $\glspl{state}_{t+N+1|t+1} \in \gls{terminalSet}$.
	To conclude, the subsequent state sequence \eqref{eq:subseqSeq} is always in the zero violation set  $ \gls{probStatesSet} \ominus \lr{\glspl{liftedDistMat}\circ \gls{distSet}^{\gls{Ncvpm}}} $.

	Since \gls{case1Set} includes all states where a prediction ${\glspl{states}_{t}\in \gls{probStatesSet} \ominus \lr{\glspl{liftedDistMat}\circ \gls{distSet}^{\gls{Ncvpm}}}} $ exists  and the subsequent prediction $\glspl{states}_{t+1}$ affected by a disturbance $\bs{w}_t\in\gls{distSet}$ is in the same set, we conclude that \gls{case1Set} is a robust control invariant set.  
\end{proof}

\section{Proof of Lemma~\ref{lem:stab1}}
\label{sec:appendixa}

We start with preparations for the proof.
The optimal input $\bm{u}^*_t(\bm{x}_t)$ is obtained by solving the MPC optimal control problem \eqref{eq:cvpmoptim}. 
Based on \eqref{eq:cost}, we abbreviate the stage cost and the terminal cost by ${l(\bm{x}_t,\bm{u}_t) = ||\gls{state}||_{\gls{stateCost}}^2 + ||\gls{input}||_{\gls{inputCost}}^2 }$, and ${\gls{Vf}(\gls{state}) =||\gls{state}||_{\gls{terminalCost}}^2}$.
In the following, the notation introduced in \eqref{eq:notation} is used.

\begin{proof}
	For $\glspl{state}_{t|t} \in \gls{case1Set}$ with Assumption~\ref{ass:rci}, $\gls{case1Set}$ is robust control invariant according to Lemma~\ref{lem:rci}.
	We define the Lyapunov function $V(\glspl{state}_{t|t}) = \gls{cost}(\glspl{state}_{t|t}, \gls{inputs}^*)$  based on the cost \eqref{eq:cost} with the optimal feedback law $\gls{inputs}^*(\bm{x}_t)$ obtained according to \eqref{eq:cvpmoptim} where \gls{optInputSet} is  from the \gls{c1}. As $V$ is continuous, positive definite, and radially unbounded based on Assumption~\ref{ass:cost}, $\alpha_1, \alpha_2 \in \mathcal{K}_{\infty}$ exist such that \eqref{eq:Vbound} is fulfilled \cite{Khalil2002}. Additionally, $V$ is Lipschitz continuous on $\gls{case1Set}$ as $V$ only consists of quadratic terms and $\gls{case1Set}$ is bounded.
	
	Next, we prove the descent property \eqref{eq:Vdec1}. Based on ${\glspl{inputs}^*_{t} = \lr{ \bm{u}^*_{t|t}, ..., \bm{u}^*_{t+N-1|t} } }$ we obtain
	\begin{IEEEeqnarray}{rl}
		\IEEEyesnumber \label{eq:Jx0opt}
		V(\glspl{state}_{t|t})  
		&=  l\lr{\glspl{state}_{t|t},\glspl{input}_{t|t}^*}  + q\lr{\glspl{state}_{t+1|t}^*} 
	\end{IEEEeqnarray}
	where 
	$q\lr{\glspl{state}_{t+1|t}^*} =  \sum_{k=1}^{N-1} l\lr{\glspl{state}_{t+k|t}^*,\glspl{input}_{t+k|t}^*}  + \gls{Vf}(\glspl{state}^*_{t+N|t})$ 
	summarizes the total cost starting at $\glspl{state}_{t+1|t}^*$, which is used similarly in \cite{BujarbaruahEtalBorrelli2021}. 
	For $t+1$ with 
	${\glspl{inputs}^*_{t+1} }$ 
	the optimal cost is $\gls{cost}(\glspl{state}_{t+1|t+1},\glspl{inputs}_{k+1}^*)$ and with the non-optimal input sequence $(\bm{u}^*_{t+1|t}, ..., \bm{u}^*_{t+N-1|t}, \tilde{\glspl{input}})$, we obtain due to optimality an input~$ \tilde{\glspl{input}}$ satisfying Assumption~\ref{ass:cost} 
	\begin{IEEEeqnarray}{rl} \label{eq:Jx1q}
		 V(\glspl{state}_{t+1|t+1}) \IEEEyessubnumber & \leq 
		\sum_{k=1}^{\gls{Nmpc}} l\lr{\glspl{state}_{t+k|t+1},\glspl{input}_{t+k|t}^*} +  \gls{Vf}(\glspl{state}_{t+N+1|t+1}) \IEEEeqnarraynumspace\IEEEyessubnumber \\
		&= q\lr{\glspl{state}_{t+1|t+1}} \IEEEyessubnumber 
	\end{IEEEeqnarray}
	where $q\lr{\glspl{state}_{t+1|t+1}} = q \lr{\glspl{state}^*_{t+1|t} + \gls{distMat}\gls{dist}}$  according to \eqref{eq:notation2}.
	Note that at $t+1$ the optimal input at prediction step $k$ is $\glspl{input}_{t+k|t+1}^*$, based on $\glspl{state}_{t+k|t+1}$, whereas $\glspl{input}_{t+k|t}^*$, obtained at $t$, is not optimal as $\glspl{state}_{t+1|t+1}$ was affected by $\gls{dist}$.
	With \eqref{eq:Jx0opt} and \eqref{eq:Jx1q} we obtain
	 \begin{subequations}	
	\begin{align}
	&V(\glspl{state}_{t+1|t+1}) -V(\glspl{state}_{t|t})    \\
	&\leq  q\lr{\glspl{state}_{t+1|t+1}} - l\lr{\glspl{state}_{t|t}, \glspl{input}^*_{t|t}} - q\lr{\glspl{state}_{t+1|t}^*} . \label{eq:J1J0} 
	\end{align}  
	\end{subequations}
	The term $q(\cdot)$ is Lipschitz continuous due to  Assumption~\ref{ass:cost}  and the boundedness of  \gls{case1Set} and \gls{inputSet}, resulting in	
	\begin{align}
		\norm{q \lr{\glspl{state}^*_{t+1|t} + \gls{distMat}\gls{dist}} - q \lr{\glspl{state}^*_{t+1|t}}}   
		\leq L_q ||\gls{distMat}\gls{dist}||  \label{eq:q_Lipschitz}
	\end{align}
	with Lipschitz constant $L_q$. Given~\eqref{eq:J1J0} and \eqref{eq:q_Lipschitz}, 
	\begin{IEEEeqnarray}{rl}
		V(\glspl{state}_{t+1|t+1}) -V(\glspl{state}_{t|t}) 
		&\leq -l\lr{\glspl{state}_{t|t}, \glspl{input}^*_{t|t}} + L_q ||\gls{distMat}\gls{dist}|| \IEEEyessubnumber \\
		& \leq - \alpha_3(||\glspl{state}_{t|t}||)+ L_q ||\gls{distMat}\gls{dist}||. \IEEEyessubnumber \IEEEeqnarraynumspace
	\end{IEEEeqnarray}
	It is straightforward that the previous procedure holds for any $t \geq 0$. Therefore, all requirements of Definition~\ref{def:ISS} are fulfilled, i.e., $V$ is an ISS Lyapunov function and the origin of system~\eqref{eq:sys_cl} is ISS if $\glspl{state}_{t|t} \in \gls{case1Set}$.
\end{proof}

\section{Proof of Lemma~\ref{lem:stab2}}
\label{sec:appendixb}
In the following, the notation introduced in \eqref{eq:notation} is used. 
\begin{proof}
The optimal input is now determined in \eqref{eq:probOptimMalahanobis}.
We define the error between a state sequence $\gls{states}$ and the optimization variable $\gls{int_xi}$ in each time step $t+k$ as $\glspl{error}_{t+k|t}$ based on the initial value at time $t$. 
The candidate Lyapunov function is 
\begin{subequations}
\begin{align}
V'(\glspl{error}_{t|t}) &= \sum_{k=0}^{\gls{Ncvpm}-1} l'\lr{\glspl{error}_{t+k|t}^*} + \gls{Vf}'(\glspl{error}_{t+N|t}^*)  \\ \label{eq:Vcase2}
&=  l'\lr{\glspl{error}_{t|t}^*} + q'(\glspl{error}_{t+1|t}^*)
\end{align}
\end{subequations}
with an optimal input sequence $\gls{inputs}^*$, values of $\gls{int_xi}_k^*$ and stage cost ${l'\lr{\glspl{error}_{t+k|t}}  = || \glspl{error}_{t+k|t} ||_{\gls{Sigmax}^{-1}}^2 }$ according to \eqref{eq:probOptimMalahanobis}.
The terminal cost is defined as
$ {\gls{Vf}'(\glspl{error}_{t+N|t})  =\| \glspl{error}_{t+N|t}\|^2_{\bs{S}}  }$, 
based on \eqref{eq:costCase2}. 
Since $V'$ is positive definite, radially unbounded, and Lipschitz continuous on $\mathbb{R}^{\gls{dimx}}$, \eqref{eq:Vbound} holds.

Similar to Lemma~\ref{lem:stab1}, we define the total cost starting at $\glspl{error}_{t+1|t}$ as $q'(\glspl{error}_{t+1|t})$. 
For $t+1$ the optimal cost is $V'(\glspl{error}_{t+1|t+1})$ with the optimal values $\glspl{inputs}_{k+1}^*$ and  $\gls{int_xi}_{k+1}^*$. 
We apply a shifted non-optimal input sequence 
$(\glspl{input}^*_{t+1|t}, ..., \glspl{input}^*_{t+N-1|t},\bs{K}\glspl{state}^*_{t+N-1|t})$,
where the last input is determined with the feedback matrix $\bs{K}$.
Similar for the error, the shifted sequence  
$(\glspl{error}_{t+1|t}, ..., \glspl{error}_{t+N|t},(\gls{sysMat}+\gls{inputMat}\bs{K})\glspl{error}_{t+N|t})$ is used.
We obtain an upper bound
\begin{subequations}
	\begin{align}
V'(\glspl{error}_{t+1|t+1}) & \leq  \sum_{k=1}^{\gls{Ncvpm}} l'\lr{\glspl{error}_{t+k|t+1}^*} + \gls{Vf}'(\glspl{error}_{t+N+1|t+}^*)  \\
&=   q'(\glspl{error}_{t+1|t+1}^*),
\end{align}
\end{subequations}
which can be summarized to $  q'(\glspl{error}_{t+1|t+1})$ due to \eqref{eq:costCase2}.
Since $q'$ is Lipschitz continuous on $\mathbb{R}^{\gls{dimx}}$, the same arguments as in Lemma~\ref{lem:stab1} yields
\begin{IEEEeqnarray}{rl}
	V'(\glspl{error}_{t+1|t+1}) - V'(\glspl{error}_{t|t}) \leq   - \alpha_3\lr{||\glspl{error}_{t|t}^*||} + L_{q'}||\gls{distMat}\gls{dist}||  .
\IEEEyesnumber \IEEEeqnarraynumspace
\end{IEEEeqnarray}

It follows that for $\glspl{state}_{t|t} \in \mathbb{R}^{n_x}$, which includes $\glspl{state}_{t|t} \notin \gls{case1Set}$, the error $\glspl{error}_{t|t}$ 
between \eqref{eq:sys_cl}  and the optimization variable ${\gls{int_xi}\in\gls{case1Set}^{\gls{Ncvpm}}}$ is \gls{ISS}.
\end{proof}

\section*{Acknowledgment}
We thank Francesco Borrelli and Monimoy Bujarbaruah for valuable discussions.

\bibliography{./references/CVPM}
\bibliographystyle{unsrt}

\end{document}